\documentclass[12pt]{iopart}
\usepackage{graphicx}
\begin{document}
\title[Charge qubit entanglement in an array of quantum dots]{Charge qubit entanglement via
conditional single electron transfer in an array of quantum dots}

\author{A V Tsukanov}

\address{Institute of Physics and Technology, Russian
Academy of Sciences, Nakhimovsky prospect 34, Moscow 117218,
Russia} \ead{tsukanov@ftian.ru}

\begin{abstract}
We propose a novel scheme to generate entanglement among
quantum-dot-based charge qubits via sequential electron transfer
in an auxiliary quantum dot structure whose transport properties
are conditioned by qubit states. The transfer protocol requires
the utilization of resonant optical pulses combined with
appropriate voltage gate pattern. As an example illustrating the
application of this scheme, we examine the nine-qubit Shor code
state preparation together with the error syndrome measurement.

\end{abstract}
 \submitto{\JPC}
\maketitle

\maketitle

\section{Introduction}

A wide class of problems concerned with the quantum information
processing requires an efficient and robust methods of the
creation of highly-entangled states from individual qubit states
\cite{1}, \cite{2}. Therefore, the search and the development of
reliable schemes for entanglement production appears to be a
rather important quantum computational issue. Recently, several
proposals for the entanglement generation in the semiconductor
nanostructures based on the quantum dots (QDs) (see, e. g., Refs.
\cite{3} - \cite{6}) and its application (see, e. g., Refs.
\cite{7} - \cite{13}) have been made. In those works, the
fault-tolerant quantum algorithms \cite{7} - \cite{9}, the
mechanisms of the quantum teleportation \cite{10} - \cite{12}, and
the cluster state preparation in the measurement-based quantum
computations \cite{13} exploit specific entangled charge or spin
states of electrons bound in the QDs.

In particular, the logical states of an individual (physical)
qubit can be presented by two single-electron orbital states
localized in the spatially separated potential minima of a double
quantum dot (DQD) \cite{14} - \cite{22}. The quantum operations on
that so-called charge qubit can be accomplished via adiabatic
variation of the DQD confinement potential \cite{14} - \cite{17}
or by the electromagnetically induced resonant transitions between
the DQD states \cite{18} - \cite{22}. In order to perform some
non-trivial two-qubit operation on an arbitrary pair of qubits in
a quantum register one should organize the interaction between
those qubits for a finite period of time. In general, after
interaction is off, the qubits become entangled with each other,
i. e., their total wave function cannot be presented as the
product of an individual qubit wave functions. The charge qubit
entanglement via electrostatic control over the tunnel coupling
between two neighboring qubits  has been studied in Refs.
\cite{4}, \cite{6}, \cite{11}, \cite{14} - \cite{17}.

In this paper, we suggest another technique of entanglement
production in the two-dimensional array of the charge qubits. The
coupling between the qubits is now indirect and mediated by an
auxiliary structure (AS) fabricated from the quasilinear chains of
the QDs and containing a single (probe) electron in the quantized
part of its conduction band. The single-electron energy spectrum
of that structure is affected by qubit states through the
electrostatic interaction between the probe and qubit electrons.
In order to demonstrate how to entangle the qubits we analyze in
detail the probe electron dynamics in a separate quasilinear QD AS
capacitively coupled to the charge qubit. As we shall see,
appropriate variation of the confinement potential of such an AS
through the application of compensating voltages aligns the energy
levels of individual AS QDs for {\it one of two} logical qubit
states. It amounts to sharp increase of the tunnel coupling
between the AS QDs that, in its turn, allows for efficient probe
electron transfer (PET) along the AS through one of the hybridized
AS states (the transport state). The transfer protocol is
accomplished via the resonant transitions connecting two states
each localized in corresponding edge AS QD and the transport state
delocalized over the AS. The numerical simulation of electron
dynamics under tight-binding approximation confirms the
possibility of successful PET implementation.

Two points are important here. Firstly, PET from one edge AS QD to
another and back results in the implementation of the phase
operation on attached qubit. Secondly, the PET between the edge AS
QDs takes place only if the qubit is in the predetermined logical
state. Therefore, the following dynamics of probe electron
appears, in its turn, to be predetermined (conditioned) by the
qubit state. It enables us to organize conditional quantum
operations (in particular, controlled-phase operation) that is the
key part of entangled state preparation. Using these results and
assuming that PET in a complex planar AS can be divided into the
sequence of elementary PETs along its quasilinear components, it
becomes possible to construct any desired entangled state of the
charge qubits attached to that AS. Particularly, the set of
described manipulations underlying controlled-phase operations,
combined with single qubit Hadamard rotations, is sufficient for
the nine-qubit Shor encoding procedure \cite{23}. Moreover, the
conditional PET may be exploited for an error syndrome
measurements and for usual qubit state measurement as well.

The paper is organized as follows. In Section 2 we examine in
detail the energy spectrum of the auxiliary QD structure attached
to single qubit as well as the organization of conditional probe
electron dynamics in that structure. This specific operation
amounts to a single-qubit phase shift and can be used for
implementation of controlled-phase shift of target qubit in
two-qubit circuits and, therefore, for entanglement generation.
Section 3 describes the algorithm of nine-qubit Shor encoding
which exploits two-qubit gates based upon conditional probe
electron evolution to entangle the qubits. Auxiliary structure
used in the algorithm is arranged from quasilinear QD chains
considered in Section 2. We conclude our study by Section 4.

\section{The conditional electron transfer along a quasilinear chain of quantum dots }

In this Section we study a way of how to manipulate the
two-electron system composed of charge qubit and auxiliary
structure by selective driving of one of its components, namely,
of the probe electron localized in the AS. As we shall see, the
probe electron evolution in the AS depends on the qubit state,
and, in its turn, may change internal qubit phase. Both aspects
are important for an algorithm of entanglement production in
many-qubit systems.

\subsection{The model and stationary eigenstates}

Consider the single-electron AS composed of $N$ electrostatically
defined QDs $A_k$ ($k=1 - N$) that are stacked in a quasilinear
chain and separated by the finite potential barriers (figure 1).
The structure parameters are chosen in such a way that the edge QD
$A_1$ ($A_N$) contains at least two bound electron states,
namelly, the well-localized ground state $\left|
g_{1}\right\rangle$ ($\left| g_{N}\right\rangle$) and the excited
state $\left| e_{1}\right\rangle$ ($\left| e_{N}\right\rangle$)
lying closely to the barrier's top, while each internal QD $A_k$
[$k=2 - (N-1)$] contains at least one bound electron state $\left|
e_k\right\rangle$. The QDs $A_1$ and $A_N$ are supposed to be
spatially isolated from each other so that the electron tunneling
between the states $\left| g_{1}\right\rangle$ and $\left|
g_{N}\right\rangle$ is negligibly small. If an electron occupies
one of those states, it can stay there for extremely long time. We
assume that the energies $\varepsilon (e_1)$ and $\varepsilon
(e_N)$ of excited states of the edge QDs and the energies
$\varepsilon (e_k)$ of states of the internal QDs are
approximately equal to each other. Given the tunnel coupling
$\tau_k$ ($\tau_k > 0$) between the states $\left|
e_k\right\rangle$ and $\left| e_{k+1}\right\rangle$ is large as
compared with the energy difference $|\varepsilon
(e_k)-\varepsilon (e_{k+1})|$ (resonant tunneling condition),
those states become hybridized. If this condition holds for any
neighboring AS QD states $\left| e_k\right\rangle$ and $\left|
e_{k+1}\right\rangle$, then the single electron tunneling between
QDs results in formation of a $N$-fold excited state subband. Each
state in that subband is a superposition of states $\left|
e_k\right\rangle$ ($k=1 - N$) and is thus delocalized over the AS.
Here we imply that the ground states of edge QDs do not hybridize
with the excited states as long as
$\varepsilon(e_{1(N)})-\varepsilon(g_{1(N)})\gg\tau_{1(N-1)}$.

If charge qubit is positioned along the structure axis $x$, as
shown in figure 1, the energies of the AS QD levels are shifted
relative to their unperturbed values $\varepsilon (g_1)$,
$\varepsilon (g_N)$, and $\varepsilon (e_k)$ by $U_q (g_1)$, $U_q
(g_N)$, and $U_q (e_k)$, respectively, due to the
electron-electron interaction. For definiteness, those shifts are
completely associated with bare single-electron levels of the AS
QDs. The shifts depend on the logical qubit state $\left|
q\right\rangle$ ($q=0,1$) and result in the suppression of the
single electron tunneling between the neighboring AS QDs $A_k$ and
$A_{k+1}$ provided that the electrostatically induced energy
mismatch for individual QD states is larger than the corresponding
tunneling matrix element, i. e., $|U_q (e_k)-U_q
(e_{k+1})|\ge\tau_k$. In order to eliminate the electrostatic
energy shifts $U_q (e_k)$ of the exited QD levels and to recover
the resonant character of the electron tunneling along the AS, one
can apply to each AS QD $A_k$ the voltages that generate the
energy shifts $\delta U_{q} (e_k)=-U_q (e_k)$. Obviously, the sets
$\left\{\delta U_{0} (e_k)\right\}_{k=1}^N$ and $\left\{\delta
U_{1} (e_k)\right\}_{k=1}^N$ of compensating voltages for two
logical charge qubit states $\left| 0\right\rangle$ and $\left|
1\right\rangle$ are different. For example, if one uses the
voltage set corresponding to the qubit state "one" and $\left|
{[U_1 \left( {e_k } \right) - U_0 \left( {e_k } \right)]-[U_1
\left( {e_{k+1} } \right) - U_0 \left( {e_{k+1} } \right)]}\right|
\ge \tau _k$ for some $k$, the unperturbed AS excited energy
subband is reconstructed only if the qubit state is "one".
Hereafter, we shall neglect the dependency of tunneling matrix
elements $\tau_k$ on the qubit state assuming that corresponding
variations in the interdot barrier heights have small effect on
$\tau_k$. In general, one can control the values of $\tau_k$ with
the voltages on the gates defining the barriers.

We analyze our two-electron system within the tight-binding
approximation and restrict ourselves to the case that the qubit
electron is always localized in the logical subspace of the DQD.
Thus the product states $\left| {g_1 ,q} \right\rangle  = \left|
{g_1 } \right\rangle \left| q \right\rangle $, $\left| {g_N ,q}
\right\rangle  = \left| {g_N } \right\rangle \left| q
\right\rangle $, $\left| {e_1 ,q} \right\rangle  = \left| {e_1 }
\right\rangle \left| q \right\rangle $, ... , $\left| {e_N ,q}
\right\rangle  = \left| {e_N } \right\rangle \left| q
\right\rangle $ ($q$ = 0, 1)  can be used as the basis states.
 In this case, the Hamiltonian describing the stationary
two-electron states reads
\begin{equation}
H = H_{AS}  + H_q  + H_{AS - q},
\end{equation}
where $H_{AS}  = \varepsilon (g_1) \left| g_1 \right\rangle
\left\langle g_1 \right| + \varepsilon (g_N) \left|
g_N\right\rangle \left\langle g_N \right| + \sum\limits_{k = 1}^N
{\varepsilon (e_k) \left| e_k \right\rangle \left\langle
e_k\right|} -\sum\limits_{k = 1}^{N-1} {\left[\tau_k \left| e_k
\right\rangle \left\langle e_{k+1}\right| + h. c.\right]}$ is the
AS Hamiltonian, $H_q  = \varepsilon_0 \left| 0 \right\rangle
\left\langle 0 \right| + \varepsilon_1 \left| 1 \right\rangle
\left\langle 1 \right|$ is the qubit Hamiltonian, and the term
\begin{equation}
H_{AS - q}  = \sum\limits_{q = 0,\,1} {\left[ {U'_q(g_1) \left|
{g_1,q} \right\rangle \left\langle {g_1,q} \right| + U'_q(g_N)
\left| {g_N,q} \right\rangle \left\langle {g_N,q} \right| +
\sum\limits_{k = 1}^N {U'_q(e_k) \left| {e_k,q} \right\rangle }
\left\langle {e_k,q} \right|} \right]}
\end{equation}
accounts for the effective interaction between the probe and qubit
electrons. (In our model, the tunneling between the AS and the
qubit as well as the non-diagonal electrostatic terms are
completely ignored.) Here $U'_q (e_k)  = U_q (e_k) + \delta U_{q'}
(e_k)$ and we assume that $U'_q (g_1)\approx U'_q (e_1)$, $U'_q
(g_N)\approx U'_q (e_N)$. The prime at $U'_q (e_k)$ indicates on
that the external voltages generate the shifts $\delta U_{q'}
(e_k)$ compensating the electrostatic shifts provided that $q$ =
$q'$. The electrostatic coupling energies are taken in the form
\begin{equation}
U_q \left( {e_k } \right) = \frac{{U_0 }}{{1 + (1-q) L/r_0 +
\left( {N - k} \right)r_c /r_0 }},
\end{equation}
where $U_0  = {1 \mathord{\left/ {\vphantom {1 {r_0
}}}\right.\kern-\nulldelimiterspace} {r_0 }}$, $r_0$ is the
distance between the center of the AS QD $A_N$ and the nearest
minimum of the DQD corresponding to the qubit state "one", $L$ is
the distance between the potential minima of the DQD, and $r_c$ is
the distance between the centers of neighboring AS QDs. For
simplicity, the AS QDs are supposed to be equally spaced from each
other. In what follows, we shall consider the AS composed of $N$ =
20 QDs with uniform interdot tunnel coupling $\tau$ at exact
resonance when $\varepsilon(e_1)= ...=\varepsilon(e_N)$. The
dependencies of $U_q(e_k)$ on the AS QD positions are presented in
figure 2. Hereafter, we shall work with the effective atomic units
1 au = $Ry^* = m^* Ry/m_e \varepsilon ^2$ for the energy, 1 au =
$a_B^* = m_e \varepsilon a_B /m^*$ for the length, and 1 au
=$\hbar/Ry^*$ for the time, where $Ry$ is the Rydberg energy,
$a_B$ is the Bohr radius, $m_e$ is the free electron mass, $m^*$
is the effective electron mass, $\varepsilon$ is the dielectric
constant, and $\hbar$ is the Planck constant. Note that the
parameters indicated in figure 2 and used throughout the paper
correspond to realistic GaAs QD system ($Ry^*$ = 6 meV, $a_B^*$ =
10 nm).

The voltage-controlled energy shifts are parameterized by the
expression $\delta U_{q'}(e_k) = -U_{q'}(e_k) \left[ {{{\delta V }
\mathord{\left/ {\vphantom {{\delta V } {U_0 }}} \right.
\kern-\nulldelimiterspace} {U_0 }}} \right]$, so that the external
voltage compensates the electrostatic shift $U_{q'}(e_k)$ of the
energy level in the AS QD $A_k$ when $\delta V$ approaches $U_0$.
Direct numerical diagonalization of the Hamiltonian, Eq. (1),
results in two groups of two-electron eigenstates each
corresponding to the qubit's localization in one of two logical
states. The basis states $\left| {g_1 ,0} \right\rangle $, $\left|
{g_N ,0} \right\rangle $, $\left| {g_1 ,1} \right\rangle $, and
$\left| {g_N ,1} \right\rangle $ are the eigenstates of the
Hamiltonian since we neglect the tunnel coupling between the AS
ground states $\left| g_{1}\right\rangle$ and $\left|
g_{N}\right\rangle$. Their energies $E\left( {g_1 ,0} \right) =
\varepsilon \left( {g_1 } \right) + \varepsilon _0  + U'_0 \left(
{g_1 } \right)$, $E\left( {g_N ,0} \right) = \varepsilon \left(
{g_N } \right) + \varepsilon _0  + U'_0 \left( {g_N } \right)$,
$E\left( {g_1 ,1} \right) = \varepsilon \left( {g_1 } \right) +
\varepsilon _1  + U'_1 \left( {g_1 } \right)$, and $E\left( {g_N
,1} \right) = \varepsilon \left( {g_N } \right) + \varepsilon _1 +
U'_1 \left( {g_N } \right)$ depend on the qubit state, the
external voltage, and the structure geometry. The excited
eigenstates of the probe electron may be represented as the
normalized superpositions $\left| m,q \right\rangle =
\sum\limits_{k = 1}^N {C_{m,k,q} \left| {e_k ,q} \right\rangle }
$, where $m$ = 1 - $N$ and $q$ = 0, 1. Since we shall be
interested in the PET between the AS ground states $\left|
g_{1}\right\rangle$ and $\left| g_{N}\right\rangle$ localized in
the AS QDs $A_1$ and $A_N$ through the resonant temporal
population of one of the excited states $\left| m,q
\right\rangle$, it is important to know the superposition
coefficients $C_{m,1,q}$ and $C_{m,N,q}$ reflecting the weights of
that transport state in both edge AS QDs. Figures 3 (a) and 3 (b)
illustrate the dependencies of absolute values of those
coefficients for the states from the central part of excited
subband (8 $\leq m \leq$ 13, $N$ = 20) on the parameter $\delta
V/U_0$ when the compensating voltages remove the electrostatic
shifts corresponding to the qubit state "one" ($q'$ = 1). The
eigenenergies $E(m,q)$ are shown in figure 3 (c). All structure
parameters are those given in Fig. 2. As it is observed from
figures. 3 (a - c), the interaction of the AS with the qubit does
not change considerably the resonant character of single-electron
tunneling in the AS for both $q$ = 0 and $q$ = 1 cases for a given
set of parameters bringing about almost uniform shift of one
excited subband relative to another. The weight coefficients
converge to their unperturbed values for $q=1$ at the point
$\delta V =U_0$ where one has
$|C_{m,1,1}|=|C_{m,N,1}|=|C_{N-m+1,1,1}|=|C_{N-m+1,N,1}|$. For
$q=0$ a very close coefficient behavior is seen at $\delta V
{\approx} 0.4U_0$, however, with some deviations from the
symmetric picture demonstrated by the plots in the case of $q$ =
1. We have revealed that the states lying in the middle of the
excited subband conserve their hybridized structure in the range 0
$\leq \delta V \leq 2U_0$, whereas the edge subband states (not
shown) are very sensitive to the external voltage. In particular,
the tunneling collapse described above manifests itself in that
the edge subband state with $m$ = 1 ($m$ = $N$) splits off the
subband and transforms at quite large positive (negative) values
of $\delta V$ ($|\delta V| \geq 3 U_0$) into the isolated state of
the QD $A_N$ with $|C_{1,N,q}|{\approx}$ 1 ($|C_{N,N,q}|{\approx}$
1). The influence of the electrostatic interaction between the
probe and qubit electrons on the AS spectral properties becomes
stronger with the decrease in $r_0$ and/or with the increase in
$L$.

\subsection{The PET and qubit phase shift in three-level
approximation}

Next we demonstrate how to implement the PET between the ground
states $\left| g_{1}\right\rangle$ and $\left| g_{N}\right\rangle$
of the edge AS QDs $A_1$ and $A_N$ given that the qubit is in one
of its logical states, say, in the state $\left| 1\right\rangle$.
Furthermore, we shall show that the PET from $\left|
g_{1}\right\rangle$ to $\left| g_{N}\right\rangle$ followed by the
reversal PET from $\left| g_{N}\right\rangle$ to $\left|
g_{1}\right\rangle$ can produce the phase shift between logical
qubit states.

Initially, the probe electron resides in the ground state $\left|
g_{1}\right\rangle$ of the QD $A_1$, whereas the state $\left|
q\right\rangle = c_0\left| 0\right\rangle + c_1\left|
1\right\rangle$ of the qubit is arbitrary. Therefore, our
two-electron system is characterized by the state vector
\begin{equation}
\left| {\Psi _{i} } \right\rangle  = \left\{ {c_0 \exp \left[ { -
iE(g_1,0)t } \right]\left| 0 \right\rangle + c_1 \exp \left[ { -
iE(g_1,1)t } \right]\left| 1 \right\rangle } \right\}\left| {g_1}
\right\rangle .
\end{equation}
Note, that $U_0(e_1){\approx}U_1(e_1)$ (see figure 2) and
$E(g_1,0)-E(g_1,1){\approx}\varepsilon_0-\varepsilon_1$,
therefore, the phase difference between two state components in
Eq. (4) is mostly defined by the difference between qubit logical
state energies. When the compensating voltages $\left\{\delta
U_{1} (e_k)\right\}_{k=1}^N$ are turned on, the excited
two-electron eigenstates $ \left| m,1 \right\rangle $ ($m$ = 1 -
$N$) transform at $\delta V=U_0$ into the unperturbed eigenstates
with high transport properties. According to Refs. \cite{18} -
\cite{22}, to attain the indirect resonant PET (now conditioned by
the qubit state $\left| 1\right\rangle$) between the states
$\left| g_1\right\rangle$ and $\left| g_N\right\rangle$ localized
in the QD $A_1$ and $A_N$, one has to irradiate the AS either by a
pair of resonant laser pulses if
$\varepsilon(g_1)\ne\varepsilon(g_N)$ (asymmetrical case) or by
single pulse if $\varepsilon(g_1)\approx\varepsilon(g_N)$
(symmetrical case). Below we consider the asymmetrical case, where
the ground-state energies of edge QDs substantially differ from
each other. The pulse frequencies $\omega_0$ and $\omega_1$ match
the resonant transition frequencies $\omega\left( {g_1 ,r,1}
\right) = E(r,1)-E(g_1,1)$ and $\omega\left( {g_N ,r,1} \right) =
E(r,1)-E(g_N,1)$  between the localized states $\left| {g_1 ,1}
\right\rangle $ and $\left| {g_N ,1} \right\rangle $ and some
excited (transport) state $\left| r,1 \right\rangle $ [for
concreteness, ${\omega_0}={\omega\left( {g_1 ,r,1} \right)}$ and
${\omega_1}={\omega\left( {g_N ,r,1} \right)}$]. Further, the
pulse strengths $E_0$ and $E_1$ have to be chosen so that the
absolute values of the coupling coefficients $\lambda \left( {g_1
,r,1} \right) = {{E_0 d\left( {g_1 ,r,1} \right)}
\mathord{\left/{\vphantom {{E_0 d\left( {g_1 ,r,1} \right)} 2}}
\right.\kern-\nulldelimiterspace} 2}$ and $\lambda \left( {g_N
,r,1} \right) = {{E_1 d\left( {g_N ,r,1} \right)}
\mathord{\left/{\vphantom {{E_1 d\left( {g_N ,r,1} \right)} 2}}
\right.\kern-\nulldelimiterspace} 2}$, where $d\left( {g_{1} ,r,1}
\right)$ and $d\left( {g_{N} ,r,1} \right)$ are the corresponding
matrix elements of optical dipole transition, be equal to each
other, i. e., $|\lambda \left( {g_1 ,r,1} \right)| = |\lambda
\left( {g_N ,r,1} \right)| \equiv \lambda$. It is also important
that only the state $\left| r,1\right\rangle $ from the AS excited
subband is to be selectively populated during the pulse action. It
means that the differences $\Delta (r,r \pm 1,1) =
E(r,1)-E(r\pm1,1)$ between the energy of transport state
$\left|r,1\right\rangle $ and the energies of the nearest states
$\left|r\pm1,1\right\rangle $ must be large compared with the
coupling coefficient $\lambda$: $\left| {\Delta (r,r \pm 1,1) }
\right| \gg  \lambda$. Besides, we should prevent the population
of states from another excited subband corresponding to the qubit
state $\left| 0\right\rangle $, by careful choice of the structure
and pulse parameters (see Sec. 2.3). Finally, it is assumed that
$\left| {\varepsilon \left( {g_1 } \right) - \varepsilon \left(
{g_N } \right)} \right| \gg \lambda$ so that each pulse drives
only its own transition. (This requirement is relevant only in
asymmetrical case.) As soon as all above conditions are satisfied,
only three states, viz. $\left| {g_1 ,1} \right\rangle $, $\left|
{g_N ,1} \right\rangle $, and $\left| r,1 \right\rangle $, are
optically active, and the effective three-level Hamiltonian
describing the resonant transfer process under the rotating-wave
approximation ($\lambda\ll\omega_0,\omega_1$) takes the form
\begin{equation}
H_{RWA}  = \lambda \left( {g_1 ,r,1} \right)\left| {g_1 ,1}
\right\rangle \left\langle {r,1} \right| + \lambda \left( {g_N
,r,1} \right)\left| {g_N ,1} \right\rangle \left\langle {r,1}
\right| + h.c.
\end{equation}
In the rotating frame, the coherent evolution of the state vector
$\left| {\Psi _\Lambda  } \right\rangle  = a_{1,1} \left| {g_1 ,1}
\right\rangle  + a_{N,1} \left| {g_N ,1} \right\rangle +
\tilde{a}_{r,1} \left| {r,1} \right\rangle$, spanned by those
states, is governed by the non-stationary Schr\"odinger equation
$i{{\partial \left| {\Psi _\Lambda } \right\rangle }
\mathord{\left/ {\vphantom {{\partial \left| {\Psi _\Lambda  }
\right\rangle } {\partial t}}} \right. \kern-\nulldelimiterspace}
{\partial t}} = H_{RWA} \left| {\Psi _\Lambda  } \right\rangle $.
Let the pulses be switched on at $t$ = 0. For the initial
condition $\left| {\Psi _\Lambda  \left( 0 \right)} \right\rangle
= \left| {g_1 ,1} \right\rangle $, the solution of this equation
is well known (see, e. g., Ref. \cite{20}) and describes the
three-level Rabi oscillations:
\begin{equation}
a_{1,1}  = \cos ^2 \left( {\Omega _R t} \right),\,\,\,a_{N,1}  =-
\sin ^2 \left( {\Omega _R t} \right),\,\,\,\tilde{a}_{r,1}  =
-\frac{i}{\sqrt{2}}\sin \left( {2\Omega _R t} \right),
\end{equation}
where $\Omega_R=\lambda/\sqrt{2}$ is the Rabi frequency. Thus, the
complete PET from the ground state of edge AS QD $A_1$ to the
ground state of edge AS QD $A_N$ takes place in times
$T_n=(\pi/2+\pi n)/\Omega_R $, where $n$ = 0, 1, 2, ... In what
follows, we shall consider the shortest time $T_0=\pi/2\Omega_R$
as the PET time.

At the end of the optical transfer ($t$ = $T_0$), the state vector
of our system in the laboratory frame transforms into
\begin{equation}
\left| \Psi \right\rangle = c_0 \exp \left[ { - iE(g_1,0)T_0 }
\right]\left| 0 \right\rangle \left| {g_1} \right\rangle - c_1
\exp \left[ { - iE(g_N,1)T_0 } \right]\left| 1 \right\rangle
\left| {g_N} \right\rangle .
\end{equation}
After the time $\tau_0$ during which no pulses act on the AS, we
drive the probe electron from the state $\left| g_N\right\rangle$
back to the state $\left| g_1\right\rangle$ in the same manner and
then turn off the compensating voltages. As a result, the
component of the two-electron state corresponding to the qubit
state $\left| 1\right\rangle$ acquires an additional phase $\theta
= \delta E (T_0+\tau_0)$ [in comparison with the free evolution
case of Eq. (4)], where $\delta
E=E(g_N,1)-E(g_1,1)|_{\delta{V}=U_0}=\varepsilon \left( {g_N }
\right) - \varepsilon \left( {g_1 } \right)$, so that for
$t\ge2T_0+\tau_0$ the state vector reads
\begin{equation}
\left|{\Psi _f } \right\rangle  = \left\{ {c_0 \exp \left[ { -
iE(g_1,0) t} \right]\left| 0 \right\rangle  + c_1 \exp \left( { -
i\theta } \right)\exp \left[ { - iE(g_1,1) t} \right]\left| 1
\right\rangle } \right\}\left| {g_1} \right\rangle.
\end{equation}
For $\theta=\pi(2m+1)$, $m$ = 0, $\pm$1, $\pm$2, ... the above
operations amount to the phase shift of the qubit logical state
$\left| 1 \right\rangle$ by $\pi$ in the qubit frame - i. e.,
$\left| q \right\rangle \left| {g_1 } \right\rangle \to \left(
{Z\left| q \right\rangle } \right)\left| {g_1 } \right\rangle $,
where $Z = \left| 0 \right\rangle \left\langle 0 \right| - \left|
1 \right\rangle \left\langle 1 \right|$. It is worth noting that
given phase shift is defined by the difference between energies of
AS ground states rather than qubit states. If $\varepsilon \left(
{g_N } \right) - \varepsilon \left( {g_1 } \right) = 0$ (e. g.,
when QD $A_1$ and $A_N$ are identical), the phase difference
between degenerate states $\left| {g_1 } \right\rangle $ and
$\left| {g_N } \right\rangle $ is not accumulated and we need
another approach to produce a phase shift. For example, by varying
the compensating voltage on the QD $A_N$ (after the PET $\left|
{g_1 } \right\rangle {\rightarrow} \left| {g_N } \right\rangle $
now driven by single pulse have been completed), it is possible to
generate required phase shift $\theta  = \int\limits_0^{\tau _0 }
{\delta \varepsilon'\left( {t'} \right)dt'} $ by appropriate
choice of the voltage-controlled energy shift $\delta
\varepsilon'\left( {t'} \right)$ of the state $\left| {g_N }
\right\rangle $ and the variation time $\tau_0$. The time $\tau_0$
is thus exploited as the additional independent parameter to
control the qubit phase in both asymmetrical and symmetrical
cases.

\subsection{Numerical simulations of the PET in 2($N$
+2)-level case}

The formulas (5) and (6) describe an idealized three-level
evolution through which PET is realized with the probability
$p_{N,1}(T_0)=|a_{N,1}(T_0)|^2=1$. However, other excited states
$\left| m,1 \right\rangle$  with $m\ne r$ (especially those
nearest to the transport state) participate electron dynamics
resulting in some decrease of $p_{N,1}(T_0)$. As it was shown in
Ref. \cite{24}, the main reason for such behavior is
non-correlated off-resonant excitations of those states during the
pulse action. The PET probability $p_{N,1}(T_0)$ accounting for
these processes is now given by the expression \cite{24}
\begin{equation}
\begin{array}{l}
 p_{N,1} \left( {T_0 } \right) \approx 1 - \sum\limits_{m \ne r} {f_m } , \\
 f_m  = \left[ {{{2\lambda \left( {g_1 ,m,1} \right)} \mathord{\left/
 {\vphantom {{2\lambda \left( {g_1 ,m,1} \right)} {\Delta \left( {m,r,1} \right)}}} \right.
 \kern-\nulldelimiterspace} {\Delta \left( {m,r,1} \right)}}} \right]^2 \sin ^2 \left[ {{{\pi \Delta \left( {m,r,1} \right)} \mathord{\left/
 {\vphantom {{\pi \Delta \left( {m,r,1} \right)} {2\sqrt 2 \lambda \left( {g_1 ,r,1} \right)}}} \right.
 \kern-\nulldelimiterspace} {2\sqrt 2 \lambda \left( {g_1 ,r,1} \right)}}} \right], \\
 \end{array}
\end{equation}
where $\Delta (m,r,1)=E(m,1)-E(r,1)$ and we require that
$|\lambda(g_1,m,1)|{\approx}|\lambda(g_N,m,1)|$ for arbitrary $m$.
Moreover, despite of inefficiency of the PET between ground states
$\left| g_1 \right\rangle$ and $\left| g_N \right\rangle$ of the
edge AS QDs for two-electron state component in Eq. (4)
corresponding to the qubit state "zero", off-resonant excitations
from the ground state $\left| g_1,0 \right\rangle$ to the states
$\left| m,0 \right\rangle$ ($m$ = 1 - $N$) may also happen. It may
amount to noticeable reduction of the probability $p_{1,0}(T_0)$
of probe electron to stay in the state $\left| g_1,0
\right\rangle$ at the pulse end.

The PET optimization is thus considered as the search of the pulse
and structure parameters for which both $p_{N,1}(T_0)$ and
$p_{1,0}(T_0)$ are as large as possible while the transfer time
$T_0$ is rather short. In particular, as we have mentioned before,
it requires the careful choice of transport state $\left| r,1
\right\rangle$ that has to possess large coupling coefficient
$\lambda(g_1,r,1)$ and be well separated from neighboring states.
According to the results of the work \cite{24}, in a
quasi-one-dimensional structure the states belonging to the
central part of excited subband are the best candidates for this
purpose. If the QD number $N$ is even, they are the states with
indices $r=N/2$ or $r=N/2+1$, whereas for odd $N$ this is the
unpaired state with the index $r=(N+1)/2$.

In what follows, we consider a general AS+qubit state in
laboratory frame
\begin{equation}
\left| \Psi  \right\rangle  = \sum\limits_{q = 0,1}c_q {\left\{
{a_{1,q} e^{ - iE\left( {g_1 ,q} \right)t} \left| {g_1 ,q}
\right\rangle  + a_{N,q} e^{ - iE\left( {g_N ,q} \right)t} \left|
{g_N ,q} \right\rangle  + \sum\limits_{m = 1}^N {\tilde a_{m,q}
e^{ - iE\left( {m,q} \right)t} \left| {m,q} \right\rangle } }
\right\}}
\end{equation}
as the superposition of 2($N$+2) two-electron eigenstates at
$\delta V =U_0$, $q'$ = 1. Since we suppose that there are no
transitions between logical qubit states during the PET in the AS,
the subspaces $\left\{ {\left| {g_1 ,0} \right\rangle ,\left| {g_N
,0} \right\rangle ,\left| {m,0} \right\rangle _{m = 1}^N }
\right\}$ and $\left\{ {\left| {g_1 ,1} \right\rangle ,\left| {g_N
,1} \right\rangle ,\left| {m,1} \right\rangle _{m = 1}^N }
\right\}$, corresponding to different logical qubit states, are
decoupled from each other and can be treated separately. To study
coherent probe electron evolution in the AS under the influence of
a pair of resonant square laser pulses polarized along the AS axis
$x$, the non-stationary Schr\"odinger equation
\begin{equation}
i{{\partial {\bf{a}}_q} \mathord{\left/ {\vphantom {{\partial
{\bf{a}}} {\partial t}}} \right. \kern-\nulldelimiterspace}
{\partial t}} = {\bf{\Lambda }}(q)\left[ {\cos \left( {\omega _0
t} \right) + \cos \left( {\omega _1 t} \right)}
\right]{\bf{a}}_q,\,\,\,q=0,\,1,
\end{equation}
governing the probability amplitude vector ${\bf{a}}_q = \left(
{a_{1,q} ,a_{N,q} ,\tilde a_{1,q} ,...,\tilde a_{N,q}} \right)^T
$, is solved numerically for the initial condition ${\bf{a}}_q(0)
= \left( {1 ,0 ,0 ,...,0} \right)^T $. Here the matrix
${\bf{\Lambda }}(q)$ with the entries $\Lambda _{kk'}(q) =2
\lambda (k,k',q) \exp \left[ { - i\Delta (k,k',q) t} \right]$
characterizes the optical dipole coupling strength for each
transition between the two-electron eigenstates from given
subspace, and indices $k$ and $k'$ run over all of those states.
The intersubband coupling coefficients between the ground states
$\left| {g_1 ,q} \right\rangle ,\left| {g_N ,q} \right\rangle$ and
an arbitrary excited state $\left| {m,q} \right\rangle $ are
calculated within tight-binding model as
\begin{equation}
\lambda \left( {g_{1\left( N \right)} ,m,q} \right) =
\varepsilon_{field} d\left( {g_{1\left( N \right)} ,m,q}
\right)/2,\,\,\ d\left( {g_{1\left( N \right)} ,m,q}
\right)=C_{m,1\left( N \right),q}d_0,
\end{equation}
while the intrasubband coupling coefficients can be found from the
expressions
\begin{equation}
\begin{array}{l}
 \lambda \left( {g_{1(N)} ,g_{1(N)} ,q} \right) =  \mp \varepsilon_{field}{{\left( {N - 1} \right)r_c } \mathord{\left/
 {\vphantom {{\left( {N - 1} \right)r_c } 4}} \right.
 \kern-\nulldelimiterspace} 4}, \,\,\lambda \left( {g_1 ,g_N ,q} \right) =0, \\
 \lambda \left( {m,n,q} \right) = -\varepsilon_{field}\sum\limits_k {C_{m,k,q}^* C_{n,k,q} \left[ {{{\left( {N - 1} \right)} \mathord{\left/
 {\vphantom {{\left( {N - 1} \right)} 2}} \right.
 \kern-\nulldelimiterspace} 2} - k + 1} \right]r_c/2 },  \\
 \end{array}
\end{equation}
where $\varepsilon _{field} = {{ea_B^* E_0 } \mathord{\left/
 {\vphantom {{ea_B^* E_0 } {Ry^* }}} \right.
\kern-\nulldelimiterspace} {Ry^* }}$ is the field energy
(actually, the dimensionless field strength), $d_0 = \left\langle
g_1 \right|-x\left| e_1 \right\rangle $ is the matrix element of
optical dipole transition between the states $\left| g_1
\right\rangle$ and $\left| e_1 \right\rangle$ of isolated QD
$A_1$, and the origin is placed at the axis $x$ in the center of
the AS. Note, that in Eq. (11) we suppose the condition $E_0 =
E_1$, underlying the coupling coefficients symmetry
$|\lambda(g_1,m,1)| = |\lambda(g_N,m,1)|$, to be fulfilled. One
can derive Eqs. (12) and (13) from the optical dipole matrix
element's definition substituting in it the superpositional form
for an excited hybridized state $\left|m,q \right\rangle$. We
neglect in Eq. (12) the interdot optical dipole transitions
setting $\left\langle g_1 \right|-x\left| e_{k{\ne}1}
\right\rangle \approx$$\left\langle g_N \right|-x\left|
e_{k{\ne}N} \right\rangle \approx$ 0 and express in Eq. (13) the
level shifts as $\left\langle e_{k} \right|-x\left| e_{k}
\right\rangle= -\left[ {{{\left( {N - 1} \right)} \mathord{\left/
 {\vphantom {{\left( {N - 1} \right)} 2}} \right.
 \kern-\nulldelimiterspace} 2} - k + 1} \right]r_c$ and $\left\langle e_{k} \right|-x\left|
e_{k'{\ne}k} \right\rangle \approx$ 0 (the latter are much smaller
than those where $k$ = $k'$ and have been discarded). Though the
intrasubband terms have large coupling values in comparison with
the intersubband ones, their contribution into the dynamics is
minor since they oscillate at very low frequencies too far from
resonance with the driving pulses. The dipole approximation used
in Eq. (5) and Eq. (11) holds if the effective AS length
$l_c=(N-1)r_c$ is much smaller than the radiation wavelength
$\lambda_w$. In our model $l_c\sim$ 10$^{-6}$ m, $\lambda_w\sim$
10$^{-4}$ m and, therefore, one has $l_c\ll \lambda_w$.

Setting $q$ = 1, we find from numerical solution of Eq. (11) the
maximal PET probability $\max(p_{N,1})=p_{N,1}(T_0)$ and
corresponding transfer time $T_0$ as functions of
$\varepsilon_{field}$ and then calculate the probability
$p_{1,0}(T_0)=|a_{1,0}(T_0)|^2$. In figure 4 we show the
dependency of $\max(p_{N,1})$ (filled circles represent numerical
data and solid curve visualizes the approximation of Eq. (9)) as
well as the dependencies of $p_{1,0}(T_0)$ vs the field energy
$\varepsilon_{field}$ for three values of $r_0$: $p_{1,0}$ for
$r_0=2$, $p'_{1,0}$ for $r_0=3$, and  $p''_{1,0}$ for $r_0=4$
(solid lines with open squares). Here $d_0$ = 0.22 (1 au =
$ea_B^*$), $r$ = 10 (the index of the transport state), $\omega_0$
= 15.083, $\omega_1$ = 18.417 and the remaining parameters are the
same as above. We see that numerical and analytical data for
$\max(p_{N,1})$ correlate well with each other thus confirming our
arguments used to justify the application of Eq. (9). From the
other hand, it is rather difficult to construct a reliable
approximation for the probability $p_{1,0}$. If $L \gg r_0$, the
oscillations of this function can be satisfactory reproduced by
the off-resonant solution for two-level system involving the state
$\left| {g_1 ,0} \right\rangle$ and excited state $\left| {m ,0}
\right\rangle$ closest to the state $\left| {r ,1} \right\rangle$.
Otherwise, if $L \ll r_0$, the populations of both components of a
two-electron state demonstrate very close behavior because now it
is hard to distinguish between the qubit charge states and the
selectivity requirement is violated. We are interested, however,
in the situation where $L$ and $r_0$ are of the same order.

Comparing both plots we define the optimal values of
$\varepsilon_{field}$ as an abscises of the graph points where the
probabilities simultaneously achieve their maxima. It is directly
observed from figure 4 that it takes place in the interval around
$\varepsilon_{field}$ = 0.08 ($E_0 {\approx}$ 500 V/cm for GaAs)
where the probabilities of interest are $\max(p_{N,1})$ = 0.9989,
$p_{1,0}$ = 0.996 at $\varepsilon_{field}$ = 0.077 and
$\max(p_{N,1})$ = 0.9955, $p''_{1,0}$ = 0.9974 at
$\varepsilon_{field}$ = 0.08. The plot illustrating the transfer
time $T_0$ vs the field energy $\varepsilon_{field}$  (figure 5)
helps us to find corresponding values of $T_0$: $T_0$ = 846 and
$T_0$ = 833, respectively. For GaAs one has 1 au = 0.11 ps and the
transfer time is about 85 ps (that is by an order of magnitude
smaller than coherence time expected in GaAs nanostructures).
Note, that for field energies $\varepsilon_{field}\geq$ 0.12 the
PET probability substantially reduces.

Let us summarize the results obtained in this Section. We have
observed that conditional PET in a quasilinear QD chain between
the edge QDs $A_1$ and $A_N$  can be carried out with high
accuracy in relatively short times. If the probe electron is then
returned back to its initial state (viz., the ground state of the
QD $A_1$), the phase operation is performed on the qubit
positioned near QD $A_N$. From the other hand, if our quasilinear
chain is a part of complex AS (the case considered below) and the
QD $A_N$ belongs also to the adjacent AS part, the PET between
those AS parts becomes conditioned by the state of the qubit
attached to the QD $A_N$. Therefore, the probe electron can be
transferred from the QD $A_1$ via the AS  to any other qubit
provided that this first (control) qubit is in the state "one".
Thus, it makes possible to implement the controlled-phase
operation on given pair of (remote) qubits.

In general, to perform a phase shift by an angle $\theta$ on the
target qubit provided that the control qubit is in the state
"one", we have to implement conditional PET from initial QD of the
structure to the QD positioned near control qubit (if and only if
its state is "one"). Secondly, the conditional transfer from this
near-to-control-qubit QD to the QD positioned near target qubit
(again, if and only if its state is "one") should be carried out.
Therefore, if the two-qubit state is $\left| 11 \right\rangle$,
phase is accumulated in target qubit according to the scenario
given in this Section. If the two-qubit state isn't $\left| 11
\right\rangle$, the probe electron doesn't go to target qubit and,
therefore, phase isn't accumulated. It is equivalent to the
transition from two-qubit state $\left| ab \right\rangle$ to state
$[\exp(i\theta)]^{ab}\left| ab \right\rangle$, where $a, b$ = (0,
1).

\section{Nine-qubit Shor encoding and error syndrome measurement}

It is known that any realistic quantum computational scheme should
be protected from quantum errors caused by the environmental
decoherence and the external control imperfections. For this
purpose, one may use the quantum error correction codes converting
individual qubit state into the specific entangled state of
several qubits that, after decoding, restores initial single qubit
state (see, e.g., the book of Nielsen and Chuang \cite{1} for
detail). In this Section we present the algorithm realizing the
nine-qubit quantum error correction code developed by Shor
\cite{23}.

\subsection{Auxiliary structure for Shor encoding algorithm}

The encoding procedure requires us to arrange the qubits in such a
way that the interactions between them can be switched on/off on
demand in controllable manner. Making use of the results of Sec.
2, we propose the following model of auxiliary structure to
mediate the interqubit coupling and to assist the entanglement
generation needed for the implementation of Shor encoding (see
figure 6). Such an AS incorporates three branches $T$, $T'$, and
$T''$ each represented by a T-shaped QD structure. The QD $A_1$
enters all branches as their common first QD and contains the
probe electron in its ground state $\left| {g_1 } \right\rangle $
at the beginning of encoding procedure. Consider in detail the
upper branch $T$ that is attached to the qubit $q_A$ whose state
we are going to encode. Its vertical part replicates the
quasilinear AS examined in Sec. 2. The QD $A_N$ positioned near
the central (control) qubit $q_A$ is used not only for operations
on that qubit but also for operations, conditioned by the state
"one" of the qubit $q_A$, on two side (target) qubits $q_B$ and
$q_C$ placed near the QDs $B_N$ and $C_N$. Thus the QD $A_N$
serves as the common first QD for two quasilinear substructures
with the symmetry axis perpendicular to the axis $x$. The
parameters of four QDs $A_1$, $A_N$, $B_N$, and $C_N$ are chosen
so that their ground state energies satisfy following relations:
$\varepsilon(g_1)\ne\varepsilon(g_N)\ne\varepsilon(g_{B_N})$ and
$\varepsilon(g_{B_N})=\varepsilon(g_{C_N})$. Other two branches
$T'$ and $T''$ are obtained from the branch $T$ through in-plane
clockwise rotations around the center of the QD $A_1$ by the
angles $2\pi/3$ and $4\pi/3$, respectively. We see that each
branch mediates the interaction between the probe electron and a
cluster of three spatially separated charge qubits. The tunnel
couplings between the branches are controlled by the gates $G$,
$G'$, and $G''$. When a negative voltage is applied to one of
those gates, the potential barrier separating the QD $A_1$ from
corresponding branch grows exponentially, and the electron
tunneling across the barrier quickly falls. As the calculations
show the decrease of tunneling matrix element $\tau$ between two
neighboring QDs by two orders of magnitude is enough to treat
those QDs as isolated from each other. We shall consider each
blockage gate to be either opened (no voltage) or closed (the
voltage sufficient for tunneling's suppression is turned on).

The AS parameters are taken so that to neglect the differences
between the energies of all possible charge configurations in the
cluster. Those energy differences can be minimized by appropriate
choice of the QD number and/or the interdot distances in each
branch. The cluster qubits are thus considered relative each other
regardless their spatially distributed structure and,
consequently, without any conditionality between them that might
be caused by various configurations of the electron positions in
the DQD structures. The spatial addressability of cluster qubits
(understood here as the possibility of the state rotation of a
chosen cluster qubit(s) without affecting the neighboring ones) is
achieved due to the frequency and polarization selectivity of
resonant pulses.

\subsection{Nine-qubit encoding scheme}

Let all qubits except the first qubit $q_A$ be initialized. We
shall work in the qubit reference frame so that the phase
multipliers arising from the qubit state energy differences will
be omitted. We use the notation $\left| q_Aq_Bq_C \right\rangle$ =
$\left| q_A \right\rangle$$\left| q_B \right\rangle$$\left| q_C
\right\rangle$ ($q_A,q_B,q_C$ = 0, 1) for the three-qubit basis
state of the cluster coupled with the upper QD branch $T$. Similar
notations for the three-qubit states of remaining two clusters
attached to the QD branches $T'$ and $T''$ are supplied with prime
and double prime, respectively. In order to transform the
AS+qubits state $\left| {\Psi _i } \right\rangle = \left[ {c_0
\left| {0 } \right\rangle + c_1 \left| {1 } \right\rangle }
\right]\left| {00} \right\rangle \left| {000} \right\rangle
^\prime  \left| {000} \right\rangle ^{\prime \prime } \left| {g_1
} \right\rangle $ into the encoded state
\begin{equation}
\left| {\Psi _f } \right\rangle  = \left[ {c_0 \left| 0_S
\right\rangle  + c_1 \left| 1 _S\right\rangle  } \right]\left|
{g_1 } \right\rangle ,
\end{equation}
where $\left| {0_S} \right\rangle (\left| {1_S} \right\rangle)  =
2^{ - 3/2} \left[ {\left| {000} \right\rangle \pm \left| {111}
\right\rangle } \right]$ $\left[ {\left| {000} \right\rangle
^\prime   \pm \left| {111} \right\rangle ^\prime  } \right]$
 $\left[ {\left| {000} \right\rangle ^{\prime \prime }  \pm \left|
{111} \right\rangle ^{\prime \prime } } \right]$ is the Shor code
state (the code word) \cite{23} corresponding to the single qubit
state $\left| 0 \right\rangle  $ ($\left| 1 \right\rangle  $), we
need to complete the following set of operations on the qubits and
the AS.

According to standard decomposition of the encoding procedure into
the set of single- and two-qubit operations \cite{1}, one
initially should perform two controlled-NOT (CNOT) operations
where qubit $q_A$ acts as control qubit whereas qubits $q_A'$ and
$q_A''$ are used as target qubits. Given operations entangle
different clusters. Next, the Hadamard rotations are performed on
each of those qubits. Finally, two CNOT operations are to be
organized inside each of three clusters. In this case, central
cluster qubit $q_A$ ($q_A'$, $q_A''$) functions as the control
qubit whereas corresponding side cluster qubits are target ones.
Those operations entangle the qubits belonging to given cluster.
What should we do to implement all of these steps using the AS
shown in figure 6?

\subsubsection{Entanglement between the clusters}

We perform the CNOT operation on the qubits $q_A$ and $q_A'$ in
following manner. Firstly, we have to transfer the probe electron
from the ground state $\left| g_1 \right\rangle $ of the central
AS QD $A_1$ to the ground state $\left| g_N \right\rangle $ of the
QD $A_N$ provided that the state of qubit $q_A$ is "one". For this
purpose, we create the effective quantum channel connecting the
QDs $A_1$ and $A_N$ via the probe electron tunneling along
vertical part of the branch $T$. The application of compensating
voltages to the QDs $A_k$ ($k$ = 1 - N) amounts to the formation
of the quasilinear QD structure with high transport properties, as
it was described in Sec. 2.1. Note that apart from the
electrostatic interaction between the probe electron and the
electron bound in the qubit $q_A$, Eq. 3, the energy level shifts
in the QDs $A_k$ are also affected by other qubits. It is
essential, however, that required PET is to be conditioned by the
state of the qubit $q_A$ only. To make transfer process
insensitive to total charge state of remaining eight qubits, the
maximal difference between interaction energies (here, the
difference between interaction energies of the probe electron
occupying the QD $A_N$ with those qubits all loaded either in the
state "one" or "zero") should be much smaller than the coupling
coefficient $\lambda$. In this case, the optical excitation of AS
transport state will not depend on charge configuration of the
qubits $q_B$, ... , $q''_C$. We have found that this condition is
satisfied, e. g., for $\varepsilon_{field} \sim$ 0.1 and $r_c$ = 3
when the QD number in the chain $A_N$, ..., $B_N(C_N)$ is greater
than 18 - 20. The contributions originated from those interactions
to the compensating voltages can be expressed as averaged
interaction energies for given QD.

Additionally, we close the gates $G'$ and $G''$ in order to
interrupt the tunnel coupling between the QD $A_1$ and the
branches $T'$ and $T''$. The blockage gates are used here to
minimize the number of individual QD states that participate in
the formation of excited subband. Such a reduction is accompanied
with an increase in the spacings between nearest hybridized states
and, therefore, it enhances the selectivity of the driving pulses.
The numerical calculations illustrating the dependency of the
probe electron energy spectrum on the external voltages indicate
on the dissociation of total AS spectrum into the set of $N$
states delocalized over the quasilinear chain formed by QDs $A_k$
($k$ = 1 - N) and the states pertaining to other AS QDs. Residual
tunnel coupling between QD $A_N$ and its neighbors in
substructures attached to the qubits $q_B$ and $q_C$ has little
effect on transport properties of our quasilinear AS and may be
completely ruled out by the use of blockage gates analogous to
those surrounding the QD $A_1$. In what follows, we shall not
discuss in detail the formation of a quasilinear transport channel
supposing that this task can be solved in all relevant cases.

As soon as the channel has been prepared, the PET can be attained
according to the two-pulse resonant scheme since $\varepsilon
(g_1) \ne \varepsilon(g_N)$ (see Sec. 2.2). At the end of the
pulse action one has
\begin{equation}
\left| \Psi  \right\rangle  = \left| {\Psi _0 } \right\rangle  +
c_1 \Phi \left( T_0 \right)\left| {100} \right\rangle \left| {000}
\right\rangle ^\prime  \left| {000} \right\rangle ^{\prime \prime
} \left| {g_N } \right\rangle,
\end{equation}
where the state vector component corresponding to the qubit state
"zero" is denoted by $\left| {\Psi _0 } \right\rangle  = c_0
\left| {000} \right\rangle \left| {000} \right\rangle ^\prime
\left| {000} \right\rangle ^{\prime \prime } \left| {g_1 }
\right\rangle $, and $\Phi \left( T_0 \right)$ is the phase
multiplier. This form of the state vector establishes the
conditional probe electron evolution needed to construct at
following steps the controlled-phase operations on two pairs of
qubits, ($q_A$, $q_A'$) and ($q_A$, $q_A''$), where $q_A$ plays
role of control qubit.

Secondly, the Hadamard rotation ${\rm{H}}\left( {q'_A } \right) =
{\rm{2}}^{{\rm{ - 1/2}}} \left[ {\left| 0 \right\rangle + \left| 1
\right\rangle } \right]\left\langle 0 \right| + {\rm{2}}^{{\rm{ -
1/2}}} \left[ {\left| 0 \right\rangle - \left| 1 \right\rangle }
\right]\left\langle 1 \right|$ is implemented on the qubit $q'_A$:
\begin{equation}
\left| \Psi  \right\rangle  = {\rm{H}}\left( {q'_A } \right)\left|
{\Psi _0 } \right\rangle  + c_1 \Phi \left( t
\right){\rm{2}}^{{\rm{ - 1/2}}}\left| {100} \right\rangle \left(
{\left| {000} \right\rangle ^\prime   + \left| {100} \right\rangle
^\prime  } \right)\left| {000} \right\rangle ^{\prime \prime }
\left| {g_N } \right\rangle.
\end{equation}
(The single-qubit Hadamard gate can be accomplished via optical
technique described in Ref. \cite{20}.) Further, we should perform
the phase operation $Z$ on the qubit $q'_A$ via the PET between
the ground states $\left| {g_N } \right\rangle $ and $\left| {g'_N
} \right\rangle $ of the AS QDs $A_N$ and $A'_N$, conditioned by
the qubit state $\left| q_A \right\rangle $ = $\left| 1
\right\rangle $ (namely, the controlled-phase operation), without
affecting the state component ${\rm{H}}\left( {q'_A }
\right)\left| {\Psi _0 } \right\rangle$. To do this, the gates $G$
and $G'$ are opened and two laser pulses with equal strengths and
frequencies, polarized along axes $x$ and $x'$, act upon the AS.
Optically active part of the AS is now presented by two identical
quasilinear QD chains with the edge QDs $A_N$ and $A'_N$. (Some
specific features of excited state hybridization in the simplest
planar artificial molecule composed of three disk QDs were
analyzed in Ref. \cite{19}.) The pulse frequency matches the
resonant frequency for optical transitions connecting the ground
states $\left| {g_N } \right\rangle $ and $\left| {g_N' }
\right\rangle $ with the transport state. If the ground-state
energies of QDs $A_1$ and $A_N$ ($A'_N$) are substantially
different from each other, the pulses {\it do not address} the
ground state $\left| {g_1 } \right\rangle $ of the central QD
$A_1$. From other hand, the AS QDs $A_N$ and $A'_N$ are identical,
therefore, required phase shift has to be achieved through the
voltage sweeping on the QD $A'_N$ (see Sec. 2.2). After necessary
voltage manipulations and probe electron's return into the ground
state $\left| g_N \right\rangle $, the state vector becomes
\begin{equation}
\left| \Psi  \right\rangle  = {\rm{H}}\left( {q'_A } \right)\left|
{\Psi _0 } \right\rangle  + c_1 \Phi \left( t
\right){\rm{2}}^{{\rm{ - 1/2}}}\left| {100} \right\rangle \left(
{\left| {000} \right\rangle ^\prime   - \left| {100} \right\rangle
^\prime  } \right)\left| {000} \right\rangle ^{\prime \prime }
\left| {g_N } \right\rangle.
\end{equation}
As it is clearly seen, above operations are equivalent to CNOT
operation on the pair of qubits $q_A$ and $q_A'$ followed by
Hadamard rotation of the target qubit $q_A'$.

Next we repeat above procedure for the qubit $q''_A$ obtaining
\begin{equation}
\left| \Psi  \right\rangle  = {\rm{H}}\left( {q''_A }
\right){\rm{H}}\left( {q'_A } \right)\left| {\Psi _0 }
\right\rangle  + c_1 \Phi \left( t \right)2^{-1}\left| {100}
\right\rangle \left( {\left| {000} \right\rangle ^\prime   -
\left| {100} \right\rangle ^\prime  } \right)\left( {\left| {000}
\right\rangle ^{\prime \prime }  - \left| {100} \right\rangle
^{\prime \prime } } \right)\left| {g_N } \right\rangle
\end{equation}
and then transfer the probe electron from the state $\left| g_N
\right\rangle $ back to the state $\left| g_1 \right\rangle$.
Performing the Hadamard rotation ${\rm{H}}\left( {q_A } \right)$
on the qubit $q_A$ we arrive at the state vector
\begin{equation}
\left| {\tilde \Psi } \right\rangle  = \left[ {c_0 \left| {\tilde
\Psi _0 } \right\rangle  + c_1 \Phi \left( T_1 \right)\left|
{\tilde \Psi _1 } \right\rangle } \right]\left| {g_1 }
\right\rangle,
\end{equation}
where $\left| {\tilde \Psi _{0(1)} } \right\rangle  = 2^{ -
\frac{3}{2}} \left( {\left| {000} \right\rangle  \pm \left| {100}
\right\rangle } \right)\left( {\left| {000} \right\rangle ^\prime
\pm \left| {100} \right\rangle ^\prime  } \right)\left( {\left|
{000} \right\rangle ^{\prime \prime }  \pm \left| {100}
\right\rangle ^{\prime \prime } } \right)$. Note, that at this
stage we have realized on the qubits $q_A$, $q'_A$, and $q''_A$
the three-qubit Shor encoding scheme that is able to correct
trivial phase error. It is easy to write the expression for the
time $T_1$ required for above operations as
$T_1=6T_0+3T_{Had}+2\tau_0$, where we approximately set all
transfer times to be equal to $T_0$, $T_{Had}$ is the time
reserved for Hadamard rotation, and $\tau_0$ is the voltage
sweeping time. The phase
$\theta(T_1)=[\varepsilon(g_N)-\varepsilon(g_1)](T_1-T_0)$
accumulated during the process has to be equal to 2$\pi n$ ($n$ is
integer) in order to cancel the multiplier $\Phi(T_1)=\exp \left[
{ - i\theta(T_1) } \right]$. It may be attained through the
appropriate choice of the time $T_1$.

\subsubsection{Entanglement inside the clusters}

Finishing part of encoding algorithm consists in the
implementation of CNOT operations on the target cluster qubits
$q_B$ ($q'_B$, $q''_B$) and $q_C$ ($q'_C$, $q''_C$) with the
control qubit $q_A$ ($q'_A$, $q''_A$). Since
$X={\rm{H}}Z{\rm{H}}$, we perform initially the Hadamard rotation
on one of target qubits, then carry out the phase operation $Z$,
conditioned by the control qubit state "one", on that qubit and,
as the final step, perform again the Hadamard rotation on the
target qubit. This scheme is illustrated by evolution of
three-qubit cluster state $\left| {q_A q_B q_C } \right\rangle $.
The phase shift by $\pi$ of logical state "one" of target qubit
$q_B$ ($q_C$) requires the conditional PET from the QD $A_1$ to
the QD $A_N$ and then from the QD $A_N$ to the QD $B_N$ ($C_N$)
and back. The main steps that bring about the CNOT implementation
on the qubit $q_B$ are given below:
\begin{equation}
\begin{array}{l}
 \left[ {\left| {000} \right\rangle  \pm \left| {100} \right\rangle } \right]\left| {g_1 } \right\rangle \mathop  \to \limits^{PET\,A_1  \to A_N } \left| {000} \right\rangle \left| {g_1 } \right\rangle  \pm \left| {100} \right\rangle \left| {g_N } \right\rangle \mathop  \to \limits^{{\rm{H}}(q_B )}  \\
 \left| 0 \right\rangle \left[ {\left| 0 \right\rangle  + \left| 1 \right\rangle } \right]\left| 0 \right\rangle \left| {g_1 } \right\rangle  \pm \left| 0 \right\rangle \left[ {\left| 0 \right\rangle  + \left| 1 \right\rangle } \right]\left| 0 \right\rangle \left| {g_N } \right\rangle \mathop  \to \limits^{PET\,A_N  \to B_N  \to A_N }  \\
 \left| 0 \right\rangle \left[ {\left| 0 \right\rangle  + \left| 1 \right\rangle } \right]\left| 0 \right\rangle \left| {g_1 } \right\rangle  \pm \left| 0 \right\rangle \left[ {\left| 0 \right\rangle  - \left| 1 \right\rangle } \right]\left| 0 \right\rangle \left| {g_N } \right\rangle \mathop  \to \limits^{{\rm{H}}(q_B )} \left| {000} \right\rangle \left| {g_1 } \right\rangle  \pm \left| {110} \right\rangle \left| {g_N } \right\rangle . \\
 \end{array}
\end{equation}
[Here we omit in expressions of the state vector the phase
multiplier $\Phi(t)$ and normalization coefficients. The qubit
states of other two clusters remain unchanged and are not shown.]
After the repeat of CNOT on the qubit $q_C$ and the return of
probe electron into the state $\left| {g_1 } \right\rangle $, the
components of the nine-qubit state, Eq. (19), transforms into
\begin{equation}
\left| {\tilde \Psi _{0(1)} } \right\rangle  \to 2^{ -
\frac{3}{2}} \left( {\left| {000} \right\rangle  \pm \left| {111}
\right\rangle } \right)\left( {\left| {000} \right\rangle ^\prime
\pm \left| {100} \right\rangle ^\prime  } \right)\left( {\left|
{000} \right\rangle ^{\prime \prime }  \pm \left| {100}
\right\rangle ^{\prime \prime } } \right).
\end{equation}
This sequence of operations takes the time
$T_2=6T_0+4T_{Had}+2\tau_0$. Having performed the set of CNOT
operations on remaining two qubit clusters ($q_A',q_B',q_C'$) and
($q_A'',q_B'',q_C''$) we arrive at the desired encoded state, Eq.
(14). The total time $T_{tot}$ of encoding algorithm's realization
is now expressed by
\begin{equation}
T_{tot}=T_1+3T_2=24T_0+15T_{Had}+8\tau_0.
\end{equation}

\subsubsection{Time and probability estimations of encoding
procedure}

Let us evaluate the time $T_{tot}$ using the value of $T_0$
calculated in Sec. 2.3. Setting $T_{Had}{\sim}\tau_0{\sim}$ 10 ps
(the value that is longer than it was taken in Ref. \cite{20}) we
find from Eq. (22) $T_{tot}{\sim}$ 2.3 ns at $\varepsilon_{field}$
= 0.077. We see that the transfer processes are the most prolonged
part of the Shor encoding algorithm. Hence, the probability of
successful encoding implementation can be roughly estimated as
$p_{Shor}{\sim}\min[(p_{N,1})^{24}, (p_{1,0})^{24}] $ given that
all other steps were performed without errors. It yields us the
value $p_{Shor}{\sim}$ 0.91 that is, of course, insufficient for
large-scale fault-tolerant quantum computations. However, this
estimation establishes the possibility to realize given scenario
in proof-of-principle experiments. The way of further optimization
of the proposed algorithm includes the search of pulse and
structure parameters that would provide an increase of the
probability $p_{Shor}$ with simultaneous decrease of the time
$T_{tot}$.

\subsection{Error syndrome measurement}

It is known that with the help of nine-qubit Shor code one may, in
principle, correct arbitrary single-qubit quantum errors occurring
independently as long as an appropriate error syndrome measurement
strategy exists. Here we demonstrate an algorithm correcting the
$X$-type quantum error, viz., the environmentally induced qubit
state inversion. Suppose that such an error has occurred in one of
cluster qubits $q_A$, $q_B$, or $q_C$. In order to detect and
correct it, one should perform the measurements of the observables
$Z\left( {q_A } \right)Z\left( {q_B } \right) = \left[ {\left|
{00} \right\rangle \left\langle {00} \right| + \left| {11}
\right\rangle \left\langle {11} \right|} \right] \otimes I_C -
\left[ {\left| {01} \right\rangle \left\langle {01} \right| +
\left| {10} \right\rangle \left\langle {10} \right|} \right]
\otimes I_C$ and $Z\left( {q_A } \right)Z\left( {q_C } \right) =
\left[ {\left| {00} \right\rangle \left\langle {00} \right| +
\left| {11} \right\rangle \left\langle {11} \right|}
\right]\otimes I_B -  \left[ {\left| {01} \right\rangle
\left\langle {01} \right| + \left| {10} \right\rangle \left\langle
{10} \right|} \right]\otimes I_B$ (see Ref. \cite{1} for detail).
If the results of both measurements are "+1", the three-qubit
cluster state hasn't been damaged. If the measurement of $Z\left(
{q_A } \right)Z\left( {q_B } \right)$ gives "+1" ("-1") and the
measurement of $Z\left( {q_A } \right)Z\left( {q_C } \right)$
gives "-1" ("+1"), the state of the qubit $q_C$ ($q_B$) has been
inverted. Finally, if both measurements amount to the value "-1",
it indicates on the error in the qubit $q_A$. Since the corrupted
qubit is identified, one has to perform $X$ rotation on that qubit
to recover the encoded state [Eq. (14)]. The quantum phase error
$Z$ that is unitarily equivalent to the inversion error $X$ can be
corrected in the same manner. Note, that syndrome measurement
enables us to define only the damaged qubit but not the qubit
state, i. e., it conserves the entanglement.

We propose the syndrome measurement technique that is based upon
the same principle as the encoding procedure. Again, the probe
electron evolution in the AS conditioned by the states of tested
qubits helps us to reveal the quantum error. Consider the AS
composed of two identical quasilinear QD chains and used to
measure the observable $Z\left( {q_A } \right)Z\left( {q_B }
\right)$ (figure 7). The central QD $A_0$ belongs to both chains
and contains the probe electron in its ground state $\left| g_0
\right\rangle $ at the beginning of the syndrome measurement. Two
edge QDs $A_0'$ and $A_0''$ with ground states $\left| g_0'
\right\rangle $ and $\left| g_0'' \right\rangle $ are placed near
the qubits $q_A$ and $q_B$, respectively, and their single
electron states are affected by qubit states. The QD parameters
are chosen so that the energies $\varepsilon \left( {g_0 }
\right)$, $ \varepsilon \left( {g'_0 } \right)$, and $\varepsilon
\left( {g''_0 } \right)$ of QD ground states $\left| g_0
\right\rangle $, $\left| g'_0 \right\rangle $, and $\left| g''_0
\right\rangle $ are equal to each other, i. e., $\varepsilon
\left( {g_0 } \right) = \varepsilon \left( {g'_0 } \right) =
\varepsilon \left( {g''_0 } \right)$. The application of the set
of voltages, compensating the electrostatic shifts of the AS QD
energy levels when the qubits are in the state "one", results in
three different types of energy spectrum. If both qubits are in
the state "zero", an electron tunneling via the AS excited subband
remains inefficient. If the state of one of the qubits is "one",
the transport properties of the QD chain, attached to that qubit,
are recovered. Finally, if both qubits are in the state "one", the
excited states become delocalized over the whole AS. Next we
irradiate the AS by laser pulse of the frequency that matches the
resonant transition frequency for situation where {\it only one}
qubit is in its state "one". The strength and polarization of the
pulse have to be taken so that the dynamical properties of
transitions $\left| {g_0 } \right\rangle
\mathbin{\lower.3ex\hbox{$\buildrel\textstyle\rightarrow\over
{\smash{\leftarrow}\vphantom{_{\vbox to.5ex{\vss}}}}$}} \left| r
\right\rangle
\mathbin{\lower.3ex\hbox{$\buildrel\textstyle\rightarrow\over
{\smash{\leftarrow}\vphantom{_{\vbox to.5ex{\vss}}}}$}} \left|
{g'_0 } \right\rangle $
 and $\left| {g_0 } \right\rangle
\mathbin{\lower.3ex\hbox{$\buildrel\textstyle\rightarrow\over
{\smash{\leftarrow}\vphantom{_{\vbox to.5ex{\vss}}}}$}} \left| r
\right\rangle
\mathbin{\lower.3ex\hbox{$\buildrel\textstyle\rightarrow\over
{\smash{\leftarrow}\vphantom{_{\vbox to.5ex{\vss}}}}$}} \left|
{g''_0 } \right\rangle $ connecting the transport state $\left| r
\right\rangle$ and the ground states localized in the central and
edge QDs will be equivalent. It means that the transition times
and PET probabilities have to be the same for two possible charge
configurations corresponding to that case. As soon as those
requirements are met, the pulse will produce the three-level
resonant dynamics described by Eqs. (6) giving rise to the PET
between the state $\left| g_0 \right\rangle $ and {\it only one}
of states $\left| g'_0 \right\rangle $ and $\left| g''_0
\right\rangle $ (of course, if {\it only one} of qubits is in the
state "one"). Note, that in the case where both qubits are in the
state "one", despite of high degree of hybridization of the AS
excited states, the probability of probe electron excitation from
the state $\left| g_0 \right\rangle $ is very low because of the
resonant transition frequency mismatch. As a result, the probe
electron leaves the AS QD $A_0$ at the end of pulse only if the
state of tested pair of qubits is $\left| q_Aq_B \right\rangle =
\left| {01} \right\rangle $ or $\left| q_Aq_B \right\rangle =
\left| {10} \right\rangle $. Thus, an information about the
eigenvalues of syndrome operator $Z\left( {q_A } \right)Z\left(
{q_B } \right)$ can be extracted from routine detection of the
probe electron in the QD $A_0$ by measurement of current $I_{QPC}$
via the quantum point contact QPC. The eigenvalue +1 (-1) thus
corresponds to the presence (absence) of the probe electron in the
QD $A_0$. If the measurement indicates on that the probe electron
has left the QD $A_0$, the pulse is applied again to return it
into the state $\left| g_0 \right\rangle $, and the compensating
voltages are switched off. It is very important that we cannot
distinguish between two equivalent paths (otherwise the projective
measurement of qubit state would be realized). From other hand, if
one arranges driving pulse so that those transitions become
different enough (e. g., polarizing the pulse along one of the QD
branches), the setup shown in figure 7 may be used for standard
projective measurement of a charge qubit state. Other five
syndrome operators can be measured in similar way.

\section{Conclusion}

In our paper we have proposed the way of the entanglement
engineering in quantum-dot-based nanostructures. As an important
practical application of the developed framework, we have
considered (to our knowledge, for first time) the realization of
the nine-qubit Shor encoding scheme in two-dimensional array of
the semiconductor charge qubits. Highly-entangled Shor code states
protecting the quantum information against an arbitrary
single-qubit error are generated by means of appropriate optical
and voltage manipulations on the combined system involving both
qubits and the auxiliary structure. As it was shown, the PET
(conditioned by qubit state) along the auxiliary structure
together with single-qubit Hadamard rotations underly the encoding
procedure. Numerical calculations confirm the possibility of
successful implementation of the transfer protocol. In general,
the indirect qubit-qubit coupling looks as the best solution to
organize distributed quantum state manipulations in the charge
qubit array. The described principle of entanglement production
that makes use of an auxiliary system mediating the interactions
among qubits can be also adopted for quantum information
processing schemes with fully electrical control.

\newpage

\Figures

\Figure{(color online) Auxiliary structure formed by quasilinear
chain of QDs (circles) and supplied with controlling gates
(rectangles) contains a single probe electron in the ground state
$\left| g_1 \right\rangle $ of the QD $A_1$. The qubit
(single-electron double quantum dot) is positioned at the right
from the structure and loaded in the logical state "one". Below,
the AS potential profile along the structure axis $x$ is shown
schematically for the cases where the compensating voltages are
switched either on (solid line) or off (dashed line).}

\Figure{(color online) Electrostatic shifts of the AS QD levels
for two logical qubit states as a functions of QD index $k$ (QD
position). Here $U_1(e_1)-U_0(e_1)$ = 0.00082 au,
$U_1(e_N)-U_0(e_N)$ = 0.3 au, and $U_1(e_N)-U_1(e_{N-1})$ = 0.3
au.}

\Figure{(color online) Characteristics of the AS eigenspectrum vs
the control voltage parameter $\delta V/U_0$. a) The absolute
values of the weight coefficients $C_{m,1,q}$ in the QD $A_1$. b)
The absolute values of the of weight coefficients $C_{m,N,q}$ in
the QD $A_N$. c) The excited AS eigenenergies $E(m,q)$ shifted by
$\varepsilon (e_1) + \varepsilon_q$. Plots for $q$ = 1 (0) are
presented by thick solid (thin dashed) curves. The point $\delta V
= U_0$ is marked by the vertical dotted line.}

\Figure{(color online) Populations $\max(p_{N,1})$ and $p_{1,0}$,
$p'_{1,0}$, $p''_{1,0}$ vs the field energy
$\varepsilon_{field}$.}

\Figure{(color online) Transfer time $T_0$ vs the field energy
$\varepsilon_{field}$. Thin curve with filled circles shows the
numerical data and solid curve visualizes the formula
$T_0=\pi/2\Omega_R$.}

\Figure{(color online) Schematics of the auxiliary structure used
in the Shor encoding implementation (see text).}

\Figure{(color online) Auxiliary structure proposed for error
syndrome measurement on pair of qubits $q_A$ and $q_B$ (see
text).}

\newpage
\centerline{\includegraphics[width=12cm]{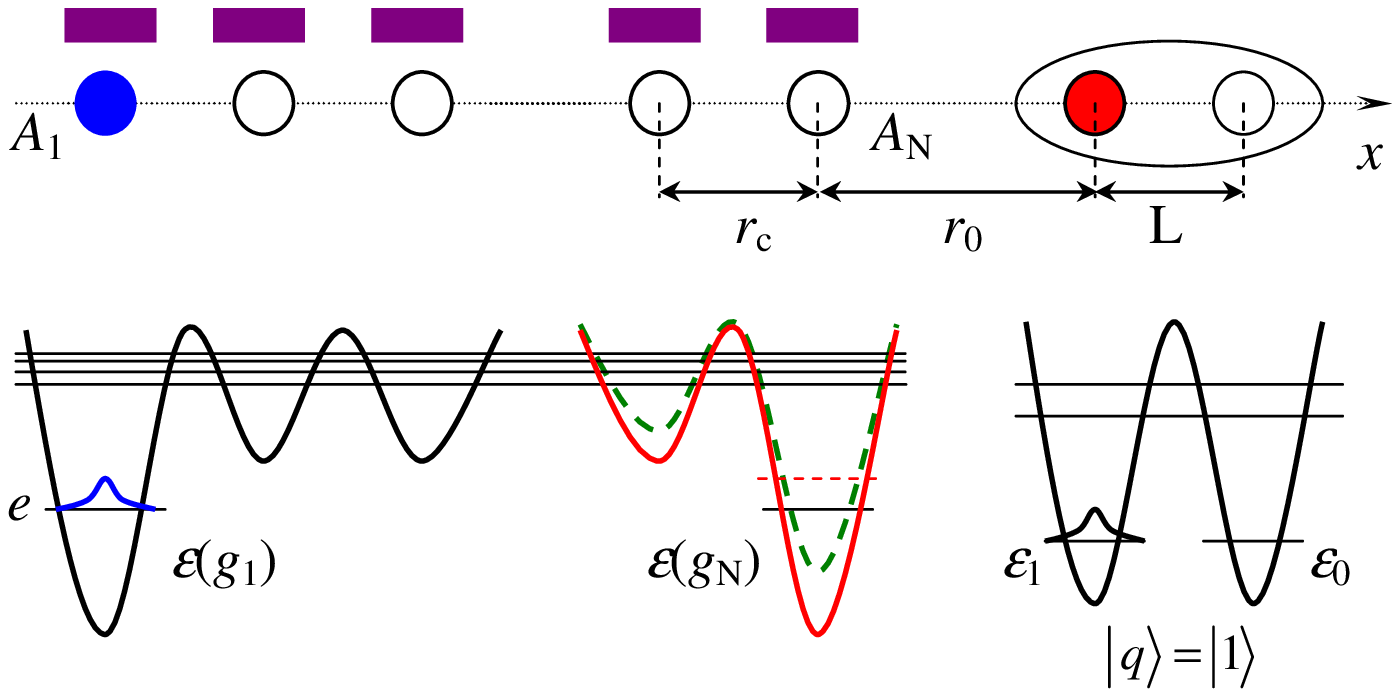}}
 \centerline{Figure 1}

\centerline{\includegraphics[width=12cm]{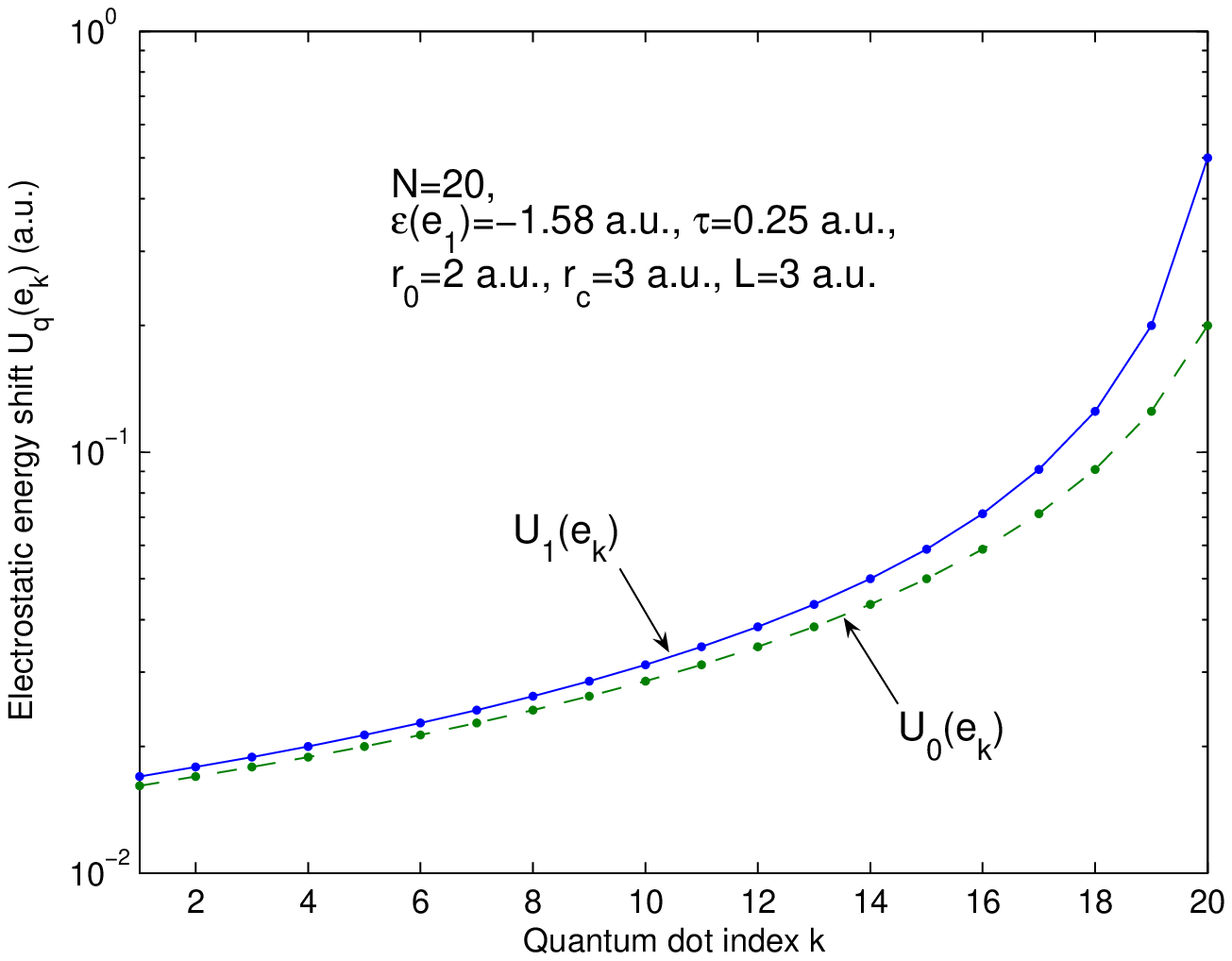}}
 \centerline{Figure 2}

\newpage
\centerline{\includegraphics[width=12cm]{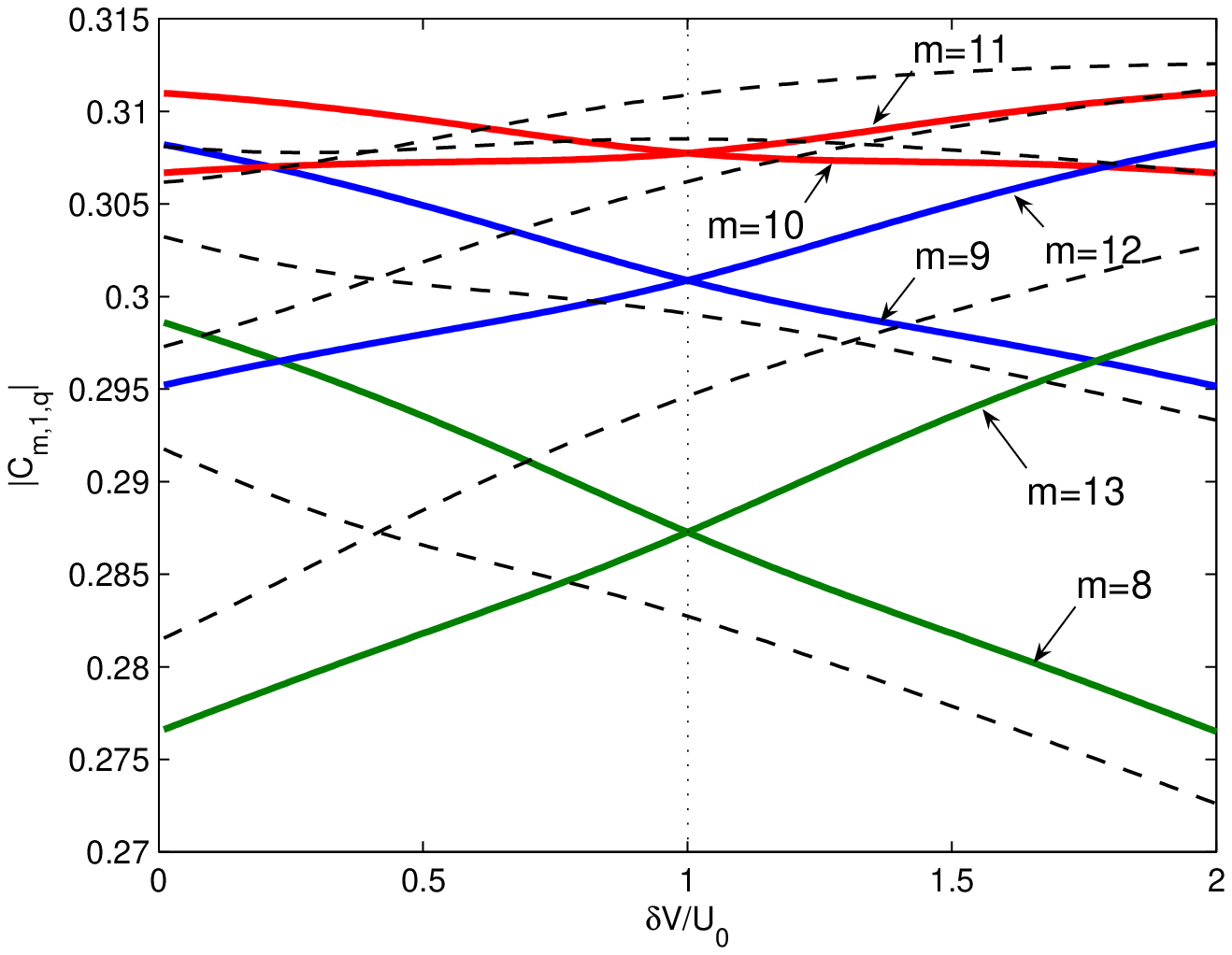}}
 \centerline{Figure 3 (a)}

\centerline{\includegraphics[width=12cm]{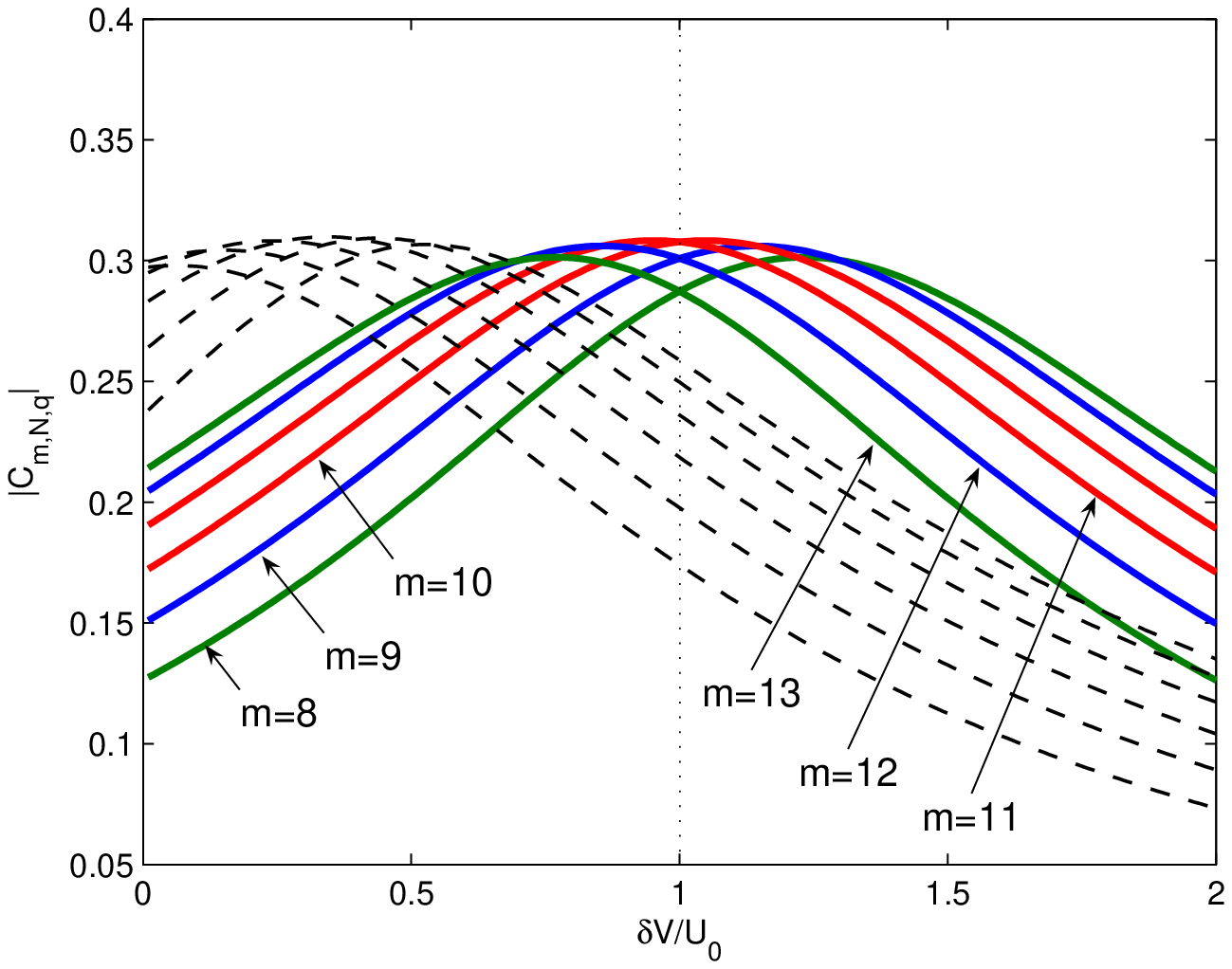}}
 \centerline{Figure 3 (b)}

\newpage
\centerline{\includegraphics[width=12cm]{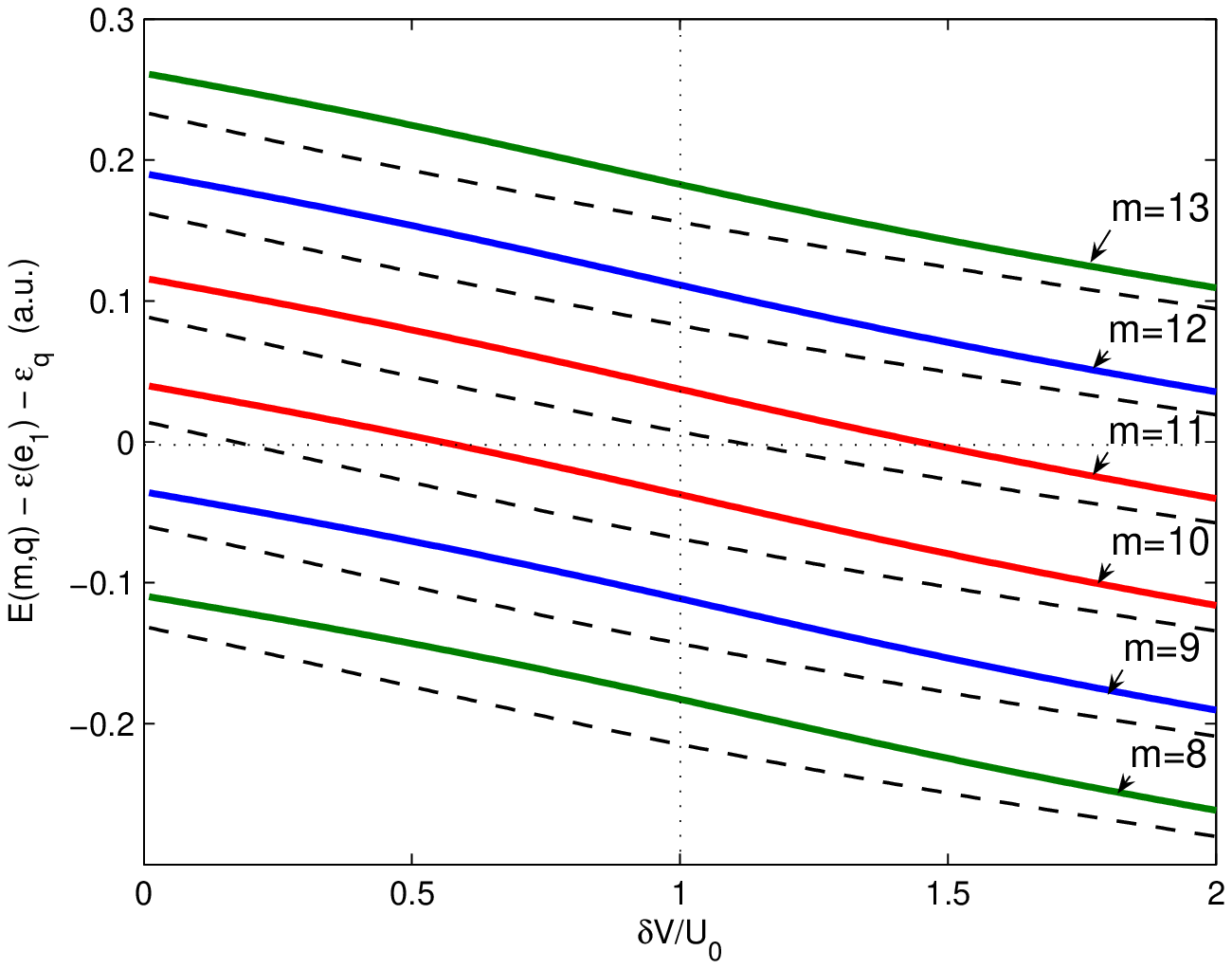}}
 \centerline{Figure 3 (c)}

\centerline{\includegraphics[width=12cm]{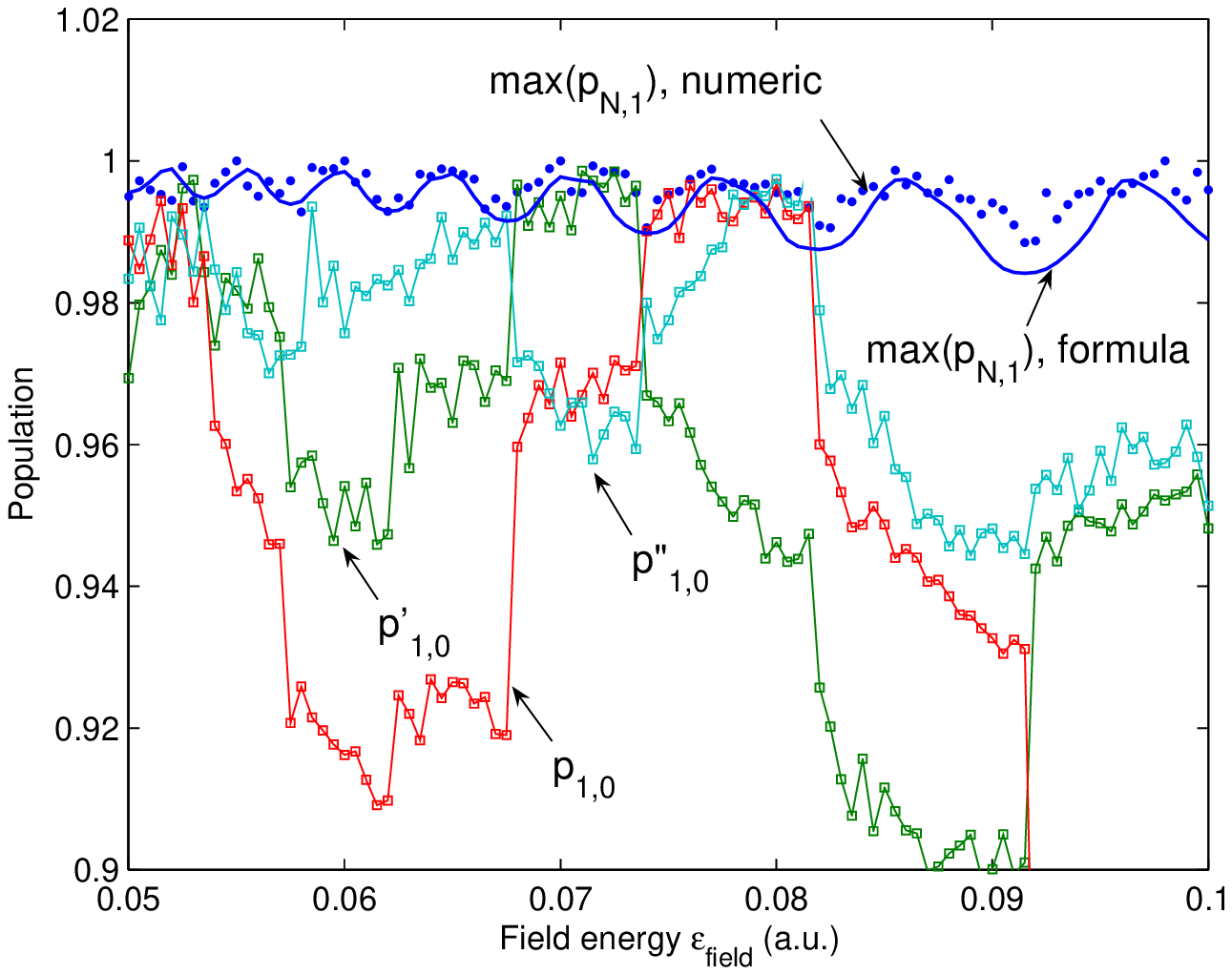}}
 \centerline{Figure 4}

\newpage
\centerline{\includegraphics[width=12cm]{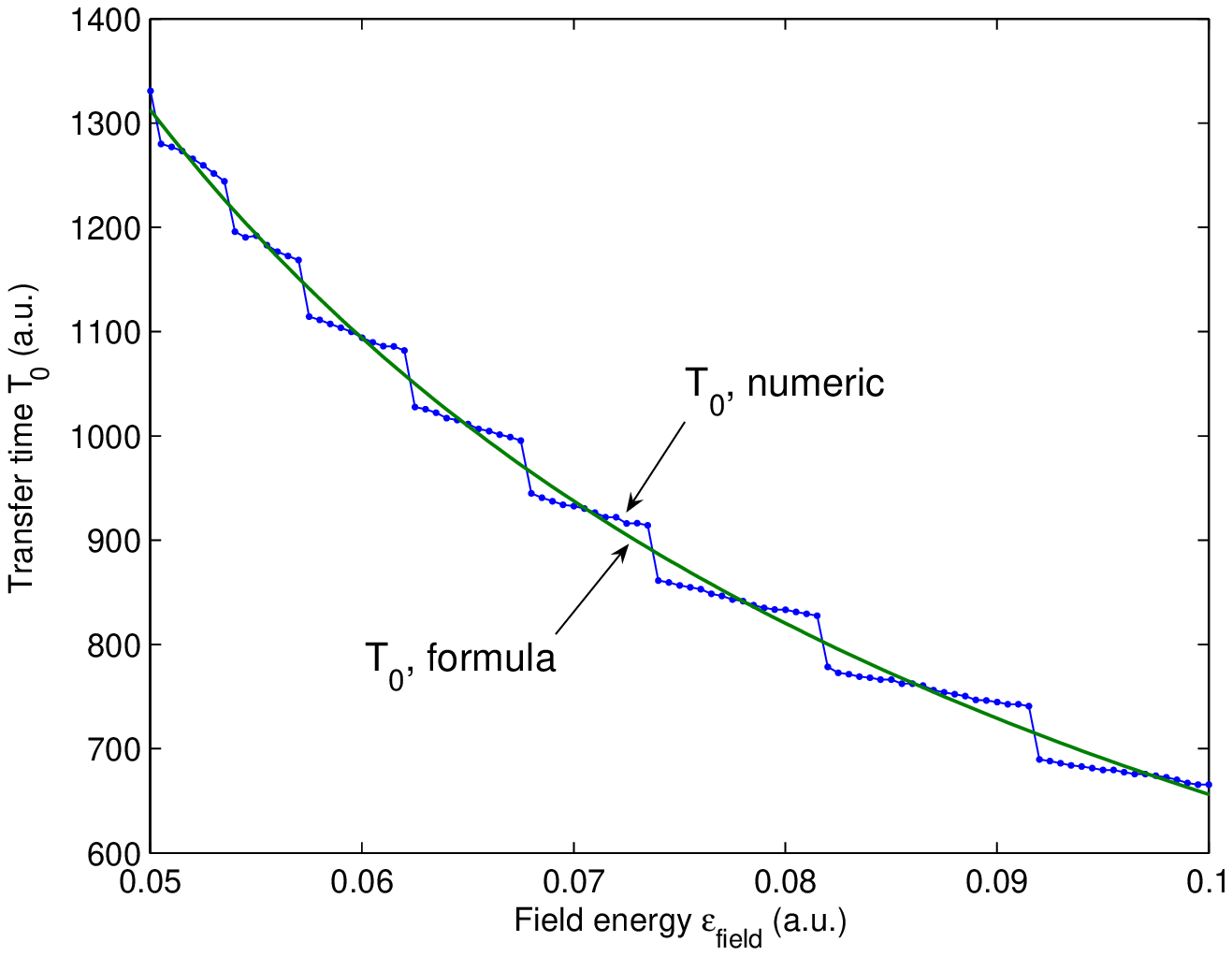}}
 \centerline{Figure 5}

\centerline{\includegraphics[width=10cm]{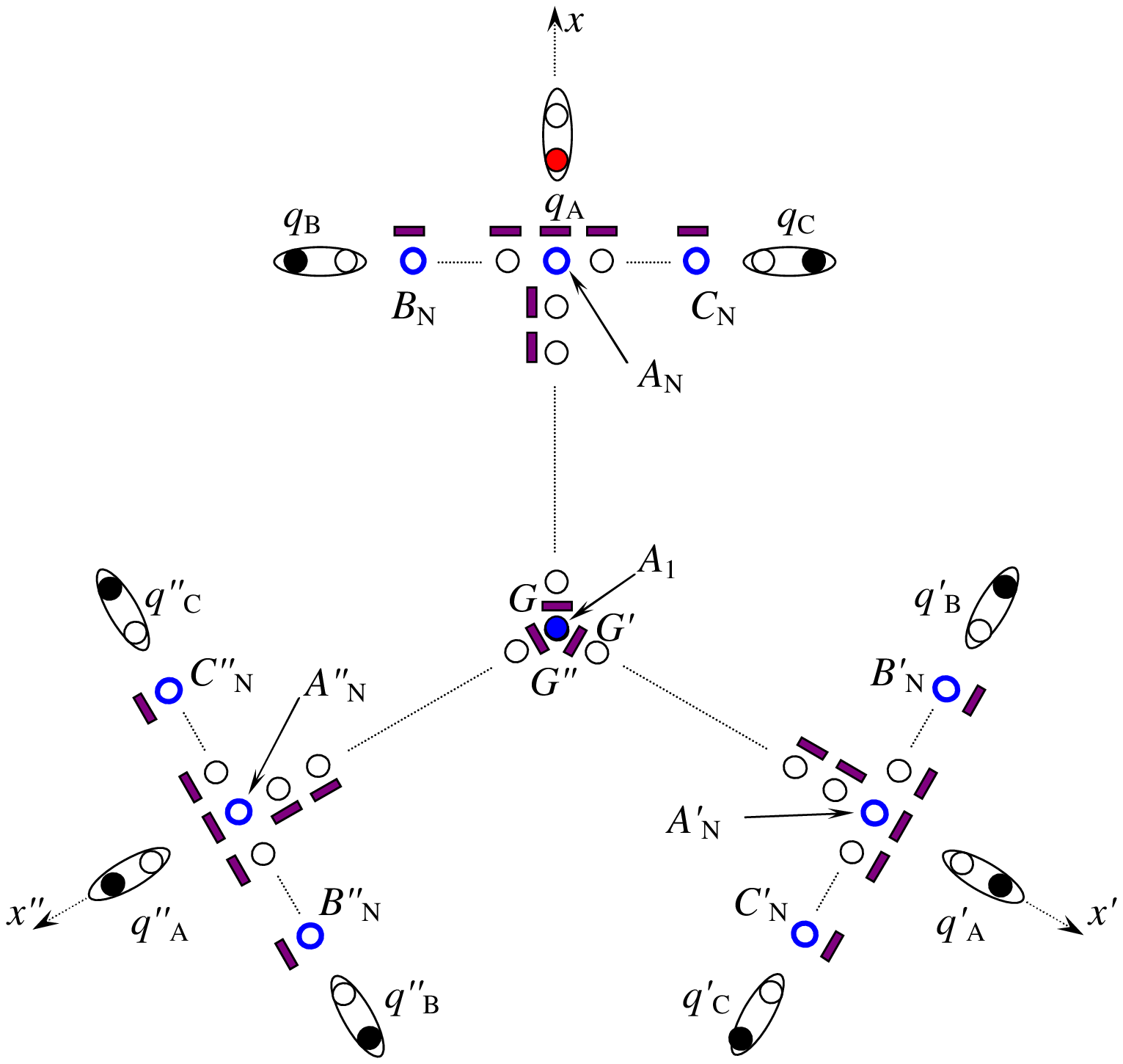}}
 \centerline{Figure 6}

\newpage
\centerline{\includegraphics[width=8cm]{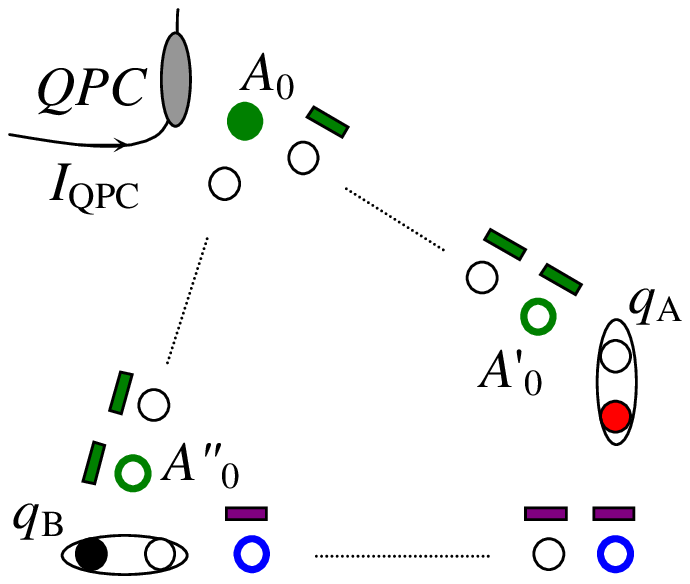}}
 \centerline{Figure 7}

\end{document}